\documentclass[twocolumn]{simplejournal}

\usepackage{booktabs}
\usepackage{algorithm}
\usepackage{algorithmicx}
\usepackage{algpseudocode}
\usepackage{setspace}
\usepackage{xcolor}
\usepackage[normalem]{ulem}

\newcommand{\AuFourCuTwo}{Au$_4$Cu$_2$}
\newcommand{\AuCuThree}{AuCu$_3$}
\newcommand{\SCAN}{r$^2$SCAN}

\title{Computing binary alloy phase diagrams with explicit configurational and vibrational entropy}

\author{
  Sarath Menon\orcidlink{0000-0002-6776-1213}$^{1,2}$\email{sarath.menon@ruhr-uni-bochum.de},
  Marvin Poul\orcidlink{0000-0002-6029-8748}$^{2}$,
  Tilmann Hickel\orcidlink{0000-0003-0698-4891}$^{3}$,
  J\"org Neugebauer\orcidlink{0000-0002-7903-2472}$^{2}$,
  Ralf Drautz\orcidlink{0000-0001-7101-8804}$^{1}$\\[6pt]
  \affil{$^1$ICAMS, Ruhr University Bochum, 44801 Bochum, Germany}\\
  \affil{$^2$Max Planck Institute for Sustainable Materials, 40237 D\"usseldorf, Germany}\\
  \affil{$^3$Bundesanstalt für Materialforschung und -prüfung, 12205 Berlin, Germany}\\
}

\date{\today}

\begin{document}

\begin{abstract}

Phase stability in multicomponent solid solutions depends on configurational entropy beyond the ideal mixing limit, but capturing it together with vibrational entropy within the same atomistic framework remains challenging.
Here, we extend non-equilibrium thermodynamic integration to composition-dependent transformations through an alchemical interpolation of the interactions, combined with Monte Carlo identity exchange moves and molecular dynamics that sample the vibrational and non-ideal configurational entropy along the integration path.
We apply the framework to the Au-Cu binary alloy using Atomic Cluster Expansion (ACE) potentials trained on density functional theory data using the LDA, PBE, and \SCAN{} functionals, and construct composition-temperature phase diagrams directly from atomistic free energies.
We find that explicit configurational sampling lowers the AuCu order-disorder transition temperature predicted by an ACE potential trained on LDA data by about 126 K, closer to the experimental value of 683~K, and substantially widens the stability range of the solid solution.
At the same time, the much larger sensitivity to the exchange-correlation functional shows that this level of agreement should not be interpreted as general predictive accuracy.
No single functional is uniformly accurate across the thermodynamic observables considered here, showing that functional sensitivity must be assessed alongside configurational effects when constructing atomistic phase diagrams.

\end{abstract}

\maketitle

\section{Introduction}

Phase diagrams are central to alloy design and materials development \cite{Chew2023}. They are typically constructed by combining experimental measurements with thermodynamic modeling, as in the CALPHAD approach \cite{Kaufman1956, Bigdeli2019}. CALPHAD has proven highly successful for many alloy systems, particularly when reliable experimental data are available. In many systems of interest, however, experimental information is sparse or unavailable, and a direct route from atomic interactions to phase stability becomes essential \cite{Chew2023}.

The direct atomistic prediction of phase stability requires the calculation of free energies as a function of temperature and composition, typically through thermodynamic integration and reversible scaling \cite{Watanabe1990, deKoning1999, Grabowski2009, Freitas2016, Menon2021}. Efficient implementation of these algorithms \cite{Menon2021}, combined with machine-learning interatomic potentials such as the Atomic Cluster Expansion (ACE) \cite{Drautz2019, Lysogorskiy2021}, now makes it feasible to cover the long time and length scales required for phase-diagram construction at near density functional theory (DFT) accuracy \cite{Menon2024}.

Extending these calculations to multicomponent systems introduces a central challenge. Alongside the vibrational contributions captured by thermodynamic integration, phase stability depends on configurational entropy associated with the distribution of atomic species over lattice sites \cite{Walsh2023, Sundman2018, Chen2021}. Lattice based methods such as the cluster variation method \cite{Sanchez1994}, cluster expansions \cite{Ozolins1998PhaseDiagrams,Ozolins1998Vibrational,vandeWalle2002, Blum2004,VanderVen2005,Jeffries2026,Sun2025}, Monte Carlo sampling \cite{Kofke1988, Kranendonk1991}, and simulations in the semi-grand canonical ensemble \cite{Sadigh2012} have established rigorous routes for treating chemical disorder and non-ideal configurational entropy. In parallel, machine learning interatomic potentials have made free energy calculations sufficiently efficient to approach phase diagram construction with near DFT accuracy. They have been used to compute anharmonic free energies, melting temperatures, and phase diagrams in automated workflows \cite{Menon2024,Poul2025,Zhu2025,Zhang2026}. For alloys, however, the construction of transferable potentials introduces an additional challenge because the relevant chemical environments vary across composition and phase space. Recent work has therefore emphasized training strategies that sample diverse chemical motifs and enable machine learning potentials to describe alloys over broad composition ranges \cite{Poul2025, Sheriff2026}. 

Most existing workflows nevertheless make at least one of three simplifications: they fix the composition and perform thermodynamic integration only at that point, they treat configurational entropy at ideal-mixing or lattice-model approximation, or they evaluate configurational and vibrational contributions in separate, decoupled steps. Computing configurational and anharmonic vibrational contributions from a single sampling, with chemical and vibrational degrees of freedom on the same footing, while resolving the composition dependence of the free energy across competing phases, remains challenging. This requires combining non-ideal configurational sampling, anharmonic vibrational contributions, and composition-dependent free energies within one efficient atomistic workflow.

A direct atomistic route is provided by combining thermodynamic integration with hybrid Monte Carlo and molecular dynamics sampling, in which configurational and vibrational degrees of freedom are treated within the same simulation \cite{Li2024}. For construction of phase diagrams, however, such approaches must remain efficient to cover the large range of compositions and temperatures spanned by competing phases. In this work, we extend non-equilibrium thermodynamic integration to composition dependent transformations through an alchemical interpolation of the interactions, combined with atomic identity exchange moves that sample the non-ideal configurational distribution along the integration path. Vibrational and configurational contributions are thereby captured within a single free energy calculation. The method is implemented in an automated workflow that, given an interatomic potential and a set of candidate phases, computes free energies over the relevant temperature composition range and constructs the corresponding phase diagram.

Beyond the sampling of configurational entropy, quantitative phase-diagram predictions also depend on the electronic-structure reference used to train the interatomic potential. Phase boundaries are controlled by small free energy differences between competing phases, and systematic errors in the underlying exchange-correlation functional can therefore translate into sizable shifts in transition temperatures and solubility limits. We therefore treat the choice of exchange-correlation functional as an explicit part of the analysis and compare phase diagrams obtained from ACE potentials trained to LDA, PBE, and \SCAN{} reference data, consistently.

We demonstrate the framework on the Au-Cu binary alloy, a system that exhibits order-disorder transitions. Interatomic interactions are described using ACE potentials trained on density functional theory data from the LDA, PBE, and \SCAN{} exchange-correlation functionals, which enables us to assess the sensitivity of the predicted phase diagram to the underlying electronic structure approximation. Using the ACE potential trained on LDA data, we show that explicit configurational sampling lowers the AuCu order-disorder transition temperature by approximately 126 K relative to the ideal-mixing approximation, moving it toward the experimental transition temperature of $\sim$683~K \cite{Sundman1998}, and substantially widens the temperature range over which the solid solution is stable, as well as the calculated solubility ranges. 
The inclusion of configurational entropy brings the ACE-LDA order-disorder transition temperature closer to experiment, but this quantitative agreement should not be overinterpreted.
The predicted melting and ordering temperatures vary substantially across LDA, PBE, and \SCAN{}, and the relative accuracy of the functionals depends on the thermodynamic observable considered.

\section{Results}

\subsection{Potential validation}

\begin{figure*}[htp!]
    \centering
    \includegraphics[width=1.0\linewidth]{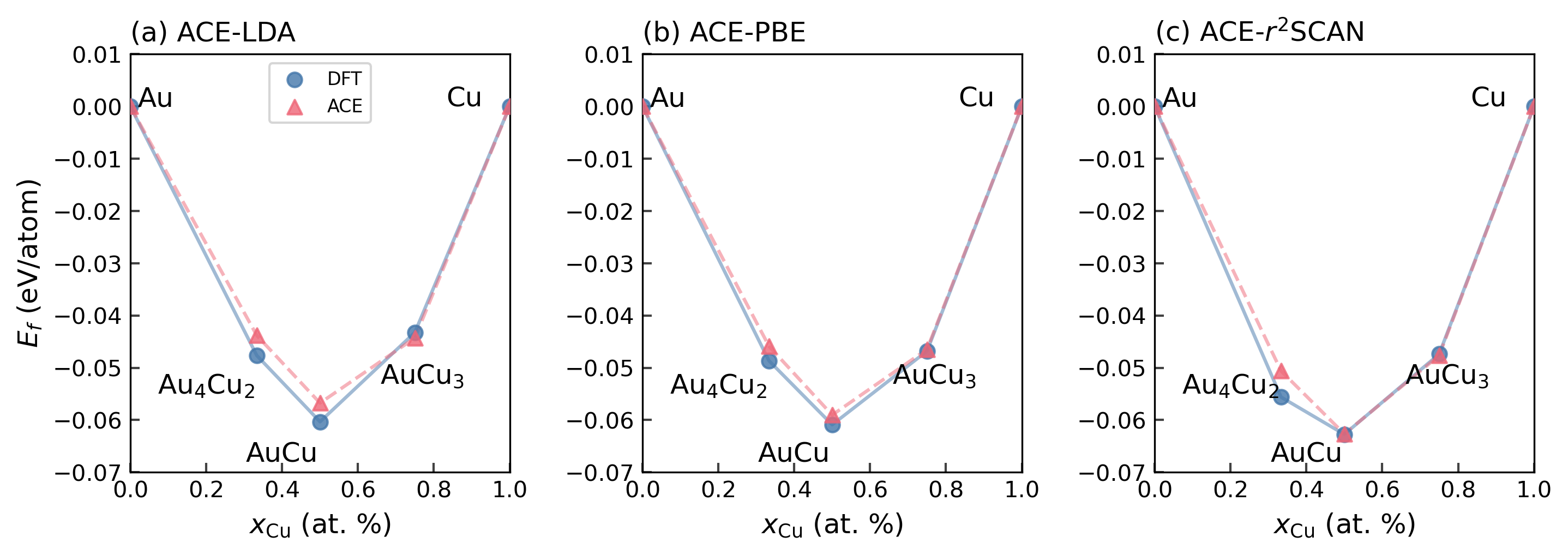}
    \caption{
    Convex hulls of the AuCu system obtained from DFT and reproduced by the corresponding ACE potentials, showing close agreement across all three functionals.
    (a) ACE-LDA (b) ACE-PBE (c) ACE-\SCAN. The compounds \AuFourCuTwo, AuCu, and \AuCuThree~appear on the convex hull for every functional, while Au$_3$Cu, which is observed in the experimental phase diagram, is not predicted to be stable by any of them.
    }
    \label{fig:convexhull}
\end{figure*}

As a first validation of the ACE potentials, we compare their predictions against the convex hulls obtained from the corresponding DFT functionals, as shown in Fig.~\ref{fig:convexhull}.
The convex hull represents the 0~K phase diagram and identifies candidate phases that may compete at finite temperature.
For all three functionals considered here, LDA, PBE, and \SCAN, the ACE potentials reproduce the phase ordering of the corresponding DFT convex hull.
Among the binary compounds, \AuFourCuTwo, AuCu, and \AuCuThree~appear on the convex hull.

A notable exception is the Au$_3$Cu phase, which appears in the experimental phase diagram but is not predicted to be stable by any of the functionals considered in this work.\cite{Zhang2014,Tian2016}
Prior work found that calculations using hybrid functionals are required to correctly describe the alloy thermodynamics in this system, at significantly increased cost.\cite{Zhang2014,Levamaki2018,Nepal2019}
In addition to the convex hull, we also evaluate further properties, including the elastic constants, energy-volume curves, and phonon dispersion relations, in order to test the transferability of the potentials beyond formation energies alone; these results are provided in the Supplementary Material.
On snapshots from molecular dynamics at high temperature, the ACE potentials reproduce the DFT free energy differences between competing phases to within a few meV/atom and the DFT forces to within a few percent (Methods Sec.~\ref{sec:fep}).

Overall, we find excellent agreement for all properties of the potentials with their respective references, despite that none of the compounds are explicitly considered during training set construction.
Details of the construction method, ASSYST, are found in sec.~\ref{sec:assyst} and Ref. \cite{Poul2025}.
Together with the phase diagrams we present below, this highlights the robustness of the ASSYST method.

\subsection{Convergence of alchemical transformations} \label{sec:convergence}

\begin{figure*}[!t]
    \centering
    \includegraphics[width=1.0\linewidth]{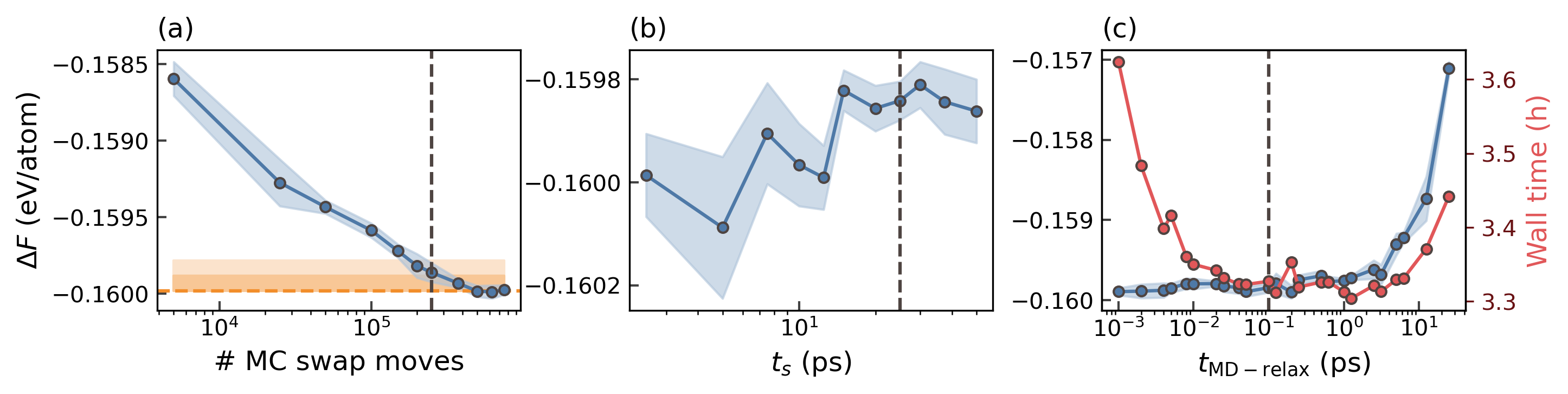}
    \caption{
    Convergence of the simulation parameters used for the alchemical free energy calculations with MC identity exchange moves.
    (a) Free-energy difference as a function of the total number of swap moves for the transformation from pure FCC Au to FCC AuCu with \(x_{\mathrm{Cu}}=0.5\). The converged free energy difference is shown in dashed dark orange line. A 0.0001 eV/atom interval is shown in dark orange, while 0.0002 eV/atom interval is in light orange.
    (b) Free-energy difference as a function of switching time for 250000 swap moves.
    (c) Free-energy difference as a function of MD relaxation time between swap blocks for 250000 swap moves and a switching time of 25 ps.
    In each panel, the selected parameter value is indicated by the gray dashed line.
    }
    \label{fig:convergence}
\end{figure*}

We next examine the convergence of the proposed methodology for calculating free energy differences as a function of composition. As a representative test case, we consider an alchemical transformation from pure Au in the FCC lattice to an FCC solid solution with \(x_{\mathrm{Cu}} = 0.5\). Calculations for the other phases are provided in Supplementary Note 9. This calculation has three independent parameters: the number of MC swap attempts $n_{\mathrm{swap}}$, the switching time $t_{\mathrm{sw}}$, and the number of MD steps $n_{\mathrm{MD}}$ between successive swap blocks.

We first evaluate the free energy difference for this transformation as a function of the number of MC identity exchange moves between Au and Cu. For this purpose, we perform between 10 and 1500 identity exchange attempts per block, with 0.1 ps of MD relaxation time between successive blocks. This corresponds to a total of 5000 to 750000 exchange attempts in each of the forward and reverse switching trajectories. The total switching time is kept fixed at 50 ps for both the forward and reverse transformations.

The resulting free energy difference for transforming half of the system to Cu is shown in Fig.~\ref{fig:convergence}(a). In the figure, the interval within 0.0001 eV/atom of the converged value is indicated in dark orange, while the interval within 0.0002 eV/atom is shown in light orange. We find that the calculated free energy enters the 0.0002~eV/atom region above $\sim$50\,000 swap attempts and the tighter 0.0001~eV/atom region above $\sim$200\,000 attempts.

Based on this analysis, we choose a total of 250000 exchange moves for the production calculations. We note that these convergence conditions are system specific, and would need tuning for a different material system. At this value, the free energy differs from the converged result by only about 0.0001 eV/atom. 
Such convergence is needed because small free energy errors shift the transition temperatures.
To first order, an error $\delta\Delta G$ in the free energy difference between two competing phases shifts their transition temperature by $\delta\Delta G/\Delta S$, where $\Delta S$ is the entropy difference between them (Supplementary Note~4).
For the AuCu order-disorder transition, $\Delta S \approx 0.5\,k_B$/atom, so that 1~meV/atom corresponds to about 23~K.
For melting of the pure elements, $\Delta S \approx 1\,k_B$/atom, and 1~meV/atom corresponds to about 12~K.
This choice provides a suitable compromise between numerical accuracy and computational cost, since further increasing the number of exchange moves leads only to marginal changes in the free energy while substantially increasing the simulation time.

With the number of swap moves fixed to the value selected above, we next examine the effect of the switching time. For this purpose, we perform calculations with switching times between 2.5 and 50 ps while keeping the total number of identity exchange attempts fixed at 250000. The resulting free energy differences are shown in Fig.~\ref{fig:convergence}(b). We find that the calculated free energy approaches convergence at around 15 ps, and we therefore choose a switching time of 25 ps for the production calculations.

Finally, we investigate the effect of the MD relaxation time between successive blocks of identity exchange attempts. To this end, we vary the number of MD steps between swap blocks from 1 to 25000, while keeping both the total number of exchange attempts and the total switching time fixed. The results are shown in Fig.~\ref{fig:convergence}(c). The free energy is insensitive to the MD relaxation time over the range $n_{\mathrm{MD}} \lesssim 1000$ steps, but begins to drift once $n_{\mathrm{MD}}$ exceeds a few thousand. In this limit, the swap attempts become too infrequent along the switching trajectory, which leads to insufficient configurational sampling and correspondingly larger errors.

At the same time, the choice of MD relaxation time has a strong effect on the computational cost. This is because each block of identity exchange attempts requires energy evaluations, whereas the energy differences for successive swap attempts can be partially reused through caching, i.e., after a successful or attempted swap only the environments of the affected atoms need to be reevaluated rather than the full system energy (details in Supplementary Note~1). As a result, grouping more swap attempts between MD segments improves computational efficiency, while for large numbers of MD steps the overall cost becomes dominated by the MD part of the simulation. Based on this analysis, we choose 100 MD steps, corresponding to 0.1 ps of relaxation time between swap blocks.

The convergence of the configurational sampling is system dependent and can vary substantially with the nature of the transformation. In particular, in systems exhibiting demixing or miscibility gaps, a significantly larger number of MC identity exchange moves may be required to obtain converged free energies. In such cases, it is often advantageous to start from a composition that is already close to the target state and for which the free energy is well converged, for example by carrying out a transformation from \(x=0.4\) to \(x=0.5\) rather than from \(x=0\) to \(x=0.5\).

Furthermore, we benchmark the alchemical transformation calculations against semi-grand canonical Monte Carlo with molecular dynamics (SGC-MC/MD) for the Au-Cu FCC solid solution at a chosen temperature, in this case, 1000 K. 
SGC-MC/MD samples the ensemble at fixed temperature, pressure, atom count, and chemical potential difference $\Delta\mu = \mu_\mathrm{Au} - \mu_\mathrm{Cu}$, allowing the Au fraction $X$ to relax to its equilibrium value at each $\Delta\mu$. The free energy of mixing is then recovered by the Legendre transformation of the semi-grand potential. The two methods agree on the free energy of mixing to within about 0.6 meV/atom on average over the entire composition range at 1000 K (see Supplementary Note~2).

In addition, this provides an opportunity to compare the computational cost of the two approaches. Both SGC-MC/MD and alchemical transformations with the same $n_{\mathrm{swap}}$ and $n_{\mathrm{MD}}$, needed at least the same simulation time $t_{\mathrm{sw}}$ for converged calculations.

The two approaches are complementary, with different practical trade-offs. Alchemical transformations require a forward and a reverse switching run at each composition ($2*t_{\mathrm{sw}}$), whereas SGC-MC/MD obtains a composition from a single run. On the other hand, the active range of the chemical potential in SGC-MC/MD is not known a priori and must be established by a preliminary coarse scan, since at sufficiently negative or positive values of $\Delta\mu$ the system collapses to a pure element state. The relation between $\Delta\mu$ and the equilibrium composition is moreover nonlinear, so an equidistant grid in $\Delta\mu$ does not yield an equidistant grid in $x$, and the chemical potentials of the production runs must be obtained by inverting the initial scan. This can be demanding in systems with narrow single-phase solubility windows, such as AuCu, where small changes in $\Delta\mu$ produce abrupt changes in $x$. Alchemical transformations, by contrast, allow the target composition to be specified directly, and each integration path remains pinned to its reference structure, whereas in SGC-MC/MD a single trajectory can spontaneously phase transform during sampling, so that identifying which phase produced a given $(\Delta\mu, x)$ point can require an explicit order-parameter analysis. The forward and reverse switching also supply a per-point uncertainty estimate from the hysteresis (See Supplementary Note 3). We note that bidirectional scans are likewise needed for SGC-MC/MD in systems with miscibility gaps or competing phases along the transformation path \cite{Mishin2004}.

\subsection{Order-disorder thermodynamics}

\begin{figure*}[!t]
    \centering
    \includegraphics[width=0.8\linewidth]{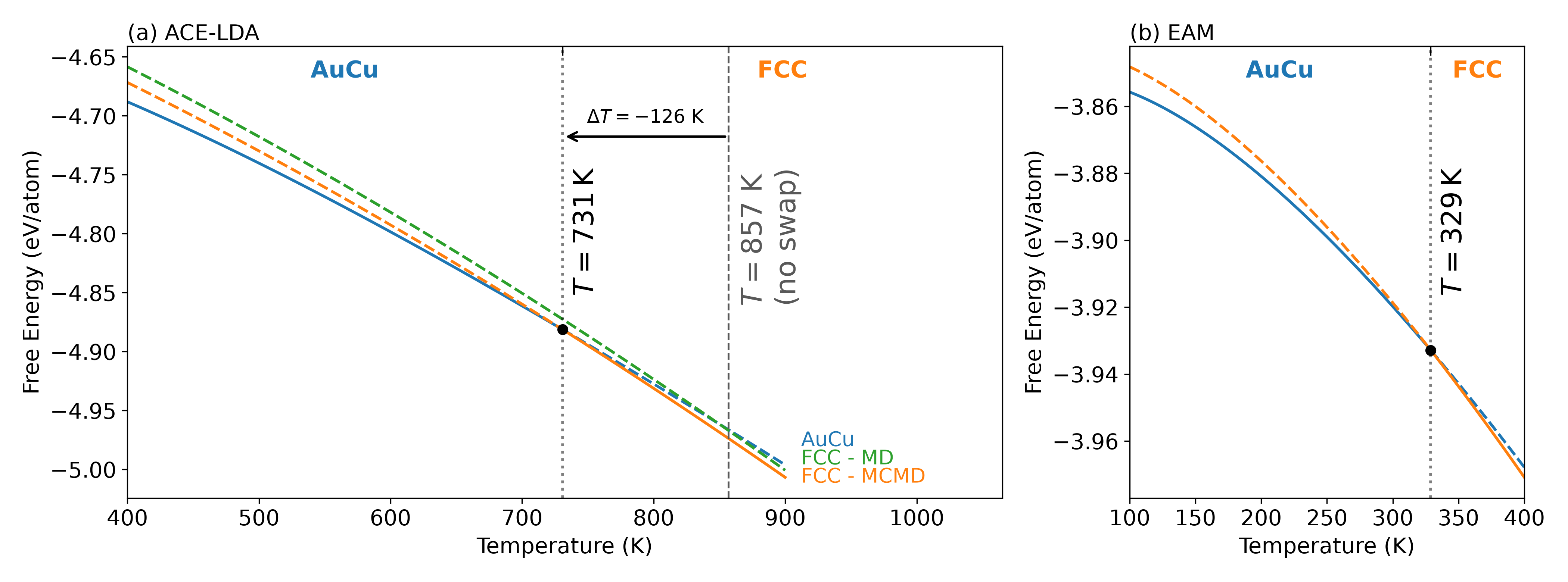}
    \caption{
    Order-disorder transition in AuCu at \(x_{\mathrm{Cu}} = 0.5\) between the disordered FCC solid solution and the ordered AuCu phase.
    (a) Free-energy curves obtained with the ACE-LDA potential, shown with and without configurational sampling through atomic identity-exchange moves. The FCC-MD curve (green) includes only the ideal configurational entropy, whereas the FCC-MCMD curve (orange) includes the configurational contribution beyond ideal mixing, obtained from Monte Carlo sampling. The intersection of the blue and green curves gives the predicted transition temperature under the ideal-mixing approximation, while the intersection of the blue and orange curves gives the transition temperature when configurational effects are accounted for.
    (b) Free-energy curves obtained with the EAM potential \cite{Gola2018}.
    The intersection of the free energy curves determines the corresponding transition temperature in each case.
    The statistical uncertainties of the free energies are below 0.5~meV/atom (Supplementary Note~4).
    }
    \label{fig:order}
\end{figure*}

Here, we examine the order-disorder transition in the AuCu system at \(x_{\mathrm{Cu}} = 0.5\), between the disordered FCC solid solution and the ordered, tetragonal (L1$_0$) AuCu phase. This transition provides a particularly useful test of the present methodology, since the relative stability of the two phases depends directly on the treatment of configurational entropy.
Experimentally, L1$_0$ AuCu transforms at about 658~K into AuCu II, a superstructure with a long period along one axis.
AuCu II disorders at about 683~K \cite{Bonneaux1999,Battezzati2001,Anraku2009}.
The CALPHAD assessment used for comparison \cite{Sundman1998} does not include AuCu II and describes the equiatomic ordered phase as L1$_0$.
We also calculated the free energy of AuCu II with ACE-LDA.
We obtained free energies 5.5--6.3~meV/atom above those of L1$_0$ AuCu between 400 and 800~K (Supplementary Note~10), so that AuCu II is not thermodynamically stabilized by ACE-LDA.
The calculated transition is therefore directly from L1$_0$ AuCu to the disordered FCC solid solution.

We first calculate the free energy curves of both phases using the ACE-LDA potential, both with and without the explicit inclusion of configurational entropy through identity exchange moves. Since the ordered AuCu and the disordered FCC solid solution are structurally distinct phases, their free energies are calculated independently, and the crossing of the two curves defines a first-order transition. When configurational entropy is not included, the predicted transition temperature is approximately 857 K. Upon including configurational entropy, the transition temperature decreases to about 731 K, a shift of 126~K relative to the ideal-mixing approximation. This shift demonstrates that explicit configurational sampling has a strong effect on the relative stability of the ordered and disordered phases, and is therefore essential for a quantitative description of the transition. The $731$~K value lies 48~K above the experimental and CALPHAD-assessed order-disorder transition temperature of $\sim$683~K \cite{Sundman1998}; however, this level of agreement should be viewed in the context of the larger functional dependence discussed below rather than as evidence of general predictive accuracy with respect to experiment. 

To test whether this agreement depends on the fitting error of the potential, we computed the difference between the ACE-LDA and DFT-LDA free energies of both phases at $x_\mathrm{Cu}=0.5$ near the transition by free energy perturbation (Methods Sec.~\ref{sec:fep}).
DFT-LDA lowers the free energy of the FCC solid solution by 1.7~meV/atom more than that of L1$_0$ AuCu.
By itself, this would lower the transition temperature, toward the experimental value.
Since the arrangement of the species is kept fixed in these calculations, they do not give the size of this shift.

The reduction in transition temperature originates from the additional stabilization of the disordered solid solution when exchange moves are included. In this case, the configurational degrees of freedom are sampled explicitly in the interacting system, so that the free energy of the solid solution is no longer described only through an idealized mixing contribution, but also reflects the actual energetics of chemically distinct local environments.

For comparison, we also calculate the transition temperature including configurational entropy using an embedded-atom method (EAM) potential \cite{Gola2018}. The EAM transition temperature is substantially lower than the experimental value, but is in close agreement with previous results obtained for the same potential using nested sampling \cite{Baldock2017}. This comparison shows that the present framework yields results consistent with other approaches that include configurational entropy contributions; the residual discrepancy between EAM and experiment originates from the limitations of the model rather than from the methodology.

We note that, in the present analysis, the \emph{stoichiometric} AuCu phase is treated as ordered and we do not explicitly account for thermal defects, that may develop within this phase at elevated temperatures. This approximation may become less accurate close to the transition, but is entirely restricted to the stoichiometric composition.  At off-stoichiometry, we include the identity exchange that sample configurational contributions. Nevertheless, the present methodology, including the temperature-scaling formalism combined with identity exchange moves, can in principle also be extended to sample disorder within the stoichiometric phase itself.
Vacancies, which may carry off-stoichiometry in other alloy systems, are neglected in this workflow.
Previous literature has found vacancies to be not important in pure Cu\,\cite{Zhu2017}, but we cannot exclude them in the intermetallic phases.
Other effects not captured by classical MLIPs are neglected as well.
This includes electronic contributions to the free energy, which have been discussed for the order-disorder transition of \AuCuThree{} \cite{paras2021contribution}.
Our free energy perturbation calculations include the electronic free energy (Methods Sec.~\ref{sec:fep}).
It changes the free energy difference between L1$_0$ AuCu and the FCC solid solution by 0.02~meV/atom and the melting temperatures by at most 3~K.

\subsection{Melting transition}

\begin{figure*}[!t]
    \centering
    \includegraphics[width=1.0\linewidth]{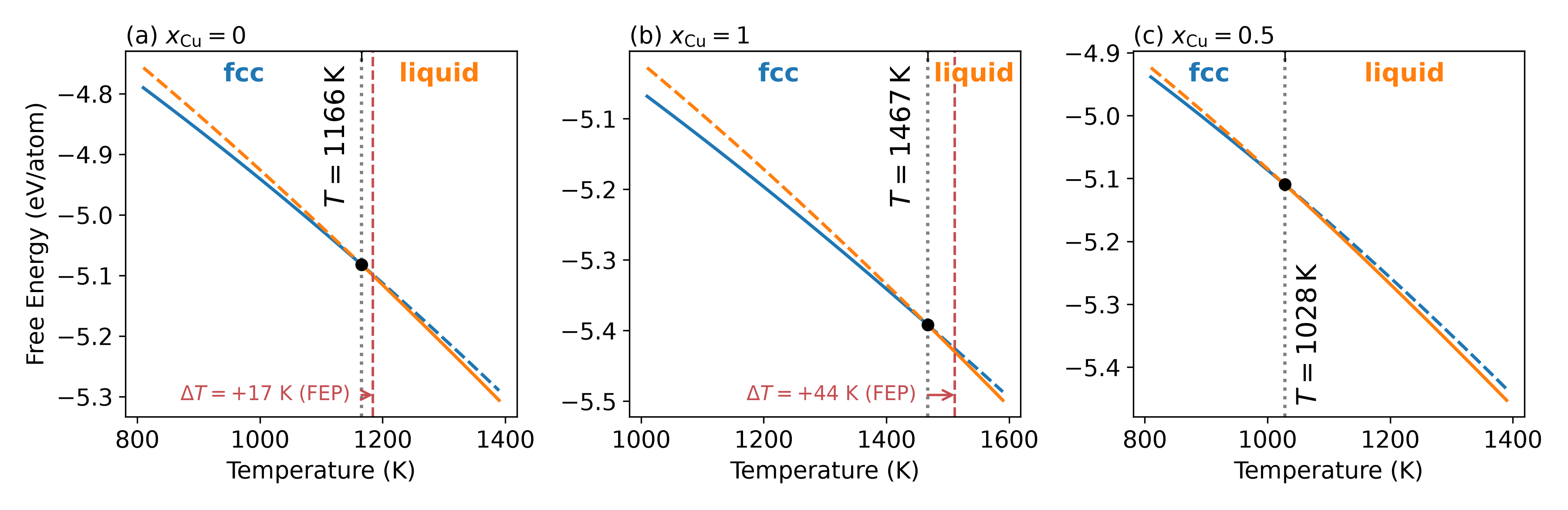}
    \caption{
    Melting temperatures of (a) pure Au, (b) pure Cu, and (c) the FCC AuCu solid solution at \(x_{\mathrm{Cu}} = 0.5\), obtained from the free energy curves of the solid and liquid phases using ACE-LDA. The intersection of the corresponding free energy curves defines the melting temperature in each case.
    In (a) and (b), red arrows mark the error of ACE-LDA with respect to DFT as calculated in Sec.~\ref{sec:fep}.
    }
    \label{fig:melting}
\end{figure*}

We next consider the melting temperatures, which define the end points of the phase diagram. These are obtained by calculating the free energies of the relevant solid phases and the liquid as a function of temperature, and then identifying the temperature at which the free energies of the two phases are equal. For pure Au, we obtain a melting temperature of 1166 K with the ACE-LDA potential, close to the DFT-LDA reference value of 1181~K \cite{Weck2020} but below the experimental melting temperature of 1337 K. The choice of exchange-correlation functional has a strong effect on the predicted melting temperature of Au. Using the ACE-PBE and ACE-\SCAN{} potentials, we obtain melting temperatures of 845 K and 886 K, respectively.
\
    Both PBE and \SCAN{} strongly underestimate the melting point compared to the experimental melting point by more than 450\,K.
    This is consistent with prior studies using MLIPs.\,\cite{shinagawa2026matlantis,andolina2024predicting}
    No DFT-PBE reference value is known to us from literature, so we have employed Free Energy Perturbation (FEP) calculations to correct for the fitting error in ACE, see again Sec.~\ref{sec:fep}.
    For PBE DFT, we obtain 861\,K including electronic contributions and 862\,K without them, for LDA DFT, 1183\,K and 1185\,K, respectively.
    Hence, the error in the prediction of ACE-LDA and ACE-PBE for Au is within 20\,K, much below the scatter between LDA and the other two functionals.

For pure Cu, the calculated melting temperatures are 1467 K, 1288 K, and 1346 K for the LDA, PBE, and \SCAN{} functionals, respectively. These values are again in good agreement with previous DFT-based estimates, which place the melting temperature at 1494 K for LDA and 1251 K for PBE \cite{Zhu2017}. Among the functionals considered here, the \SCAN{} result is closest to the experimental melting temperature of 1357 K. In contrast to Au, where LDA performs best, the Cu melting temperature is described most accurately by \SCAN{}, indicating that the ranking of the functionals is not uniform across the elemental end members.
\
    For LDA, our FEP calculations yield 1512\,K, see Sec.~\ref{sec:fep}, in agreement with \cite{Zhu2017}.

Finally, we consider the melting (or $T_0$-, as the solid solutions are melting incongruently) temperature of the FCC solid solution at \(x_{\mathrm{Cu}} = 0.5\). For this composition, we obtain $T_0$-temperatures of 1028 K for LDA, and 884 K for both PBE and \SCAN{} functionals, respectively. For the LDA model, we also compare against the no-swap calculation, in which configurational degrees of freedom are not sampled explicitly through identity exchange moves. In that case, the predicted melting temperature is 951 K. Explicit treatment of configurational entropy therefore improves the agreement with experiment (1183 K), although the effect is less pronounced than in the case of the order-disorder transition. This is expected, since at the higher temperatures relevant for melting, the ideal treatment of configurational entropy already provides a more reasonable approximation.

\subsection{Phase diagrams}

\begin{figure*}[htp!]
    \centering
    \includegraphics[width=0.8\linewidth]{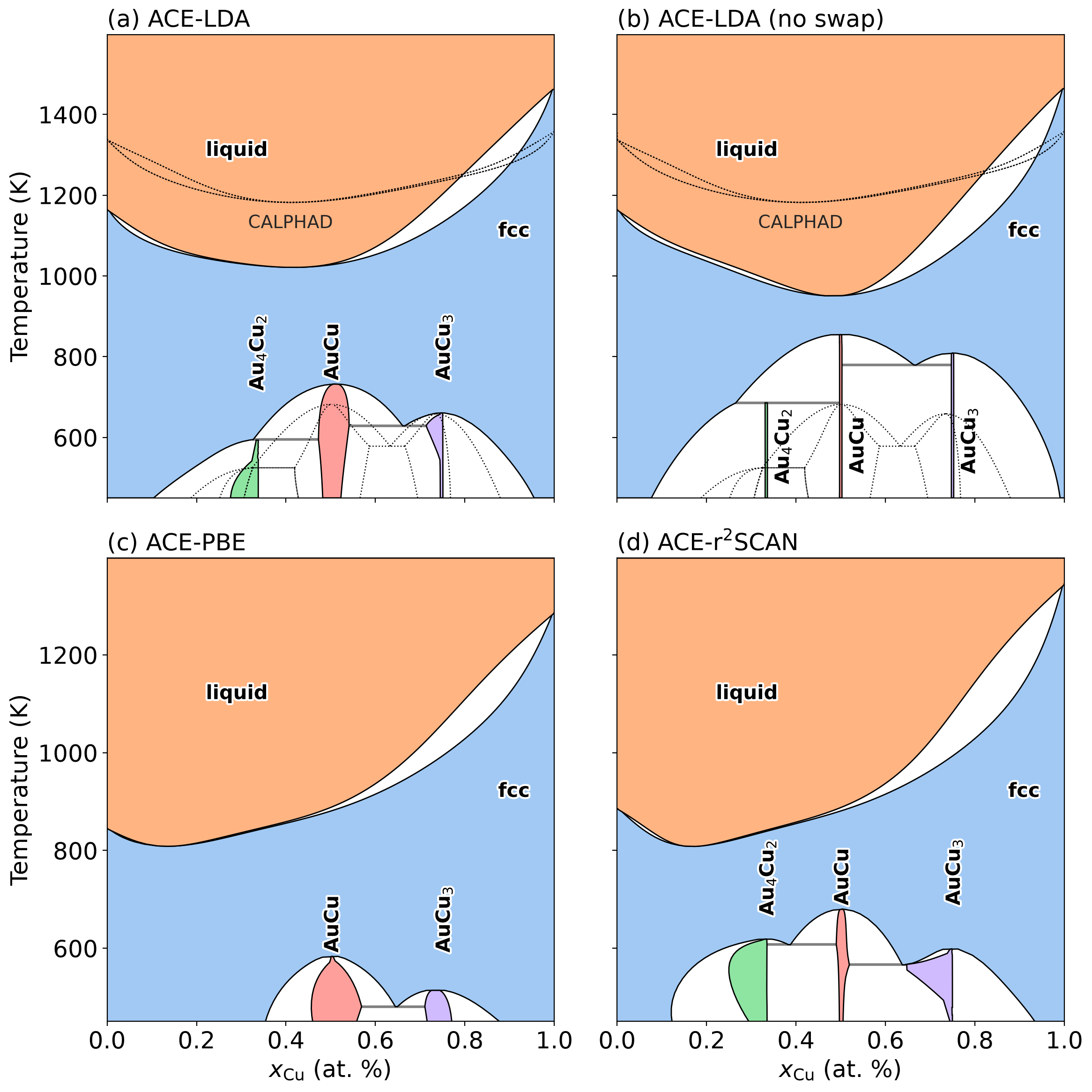}
    \caption{
    Calculated phase diagrams of AuCu obtained with (a) ACE-LDA, (b) ACE-LDA without explicit configurational sampling (no-swap; ideal-mixing entropy only, see Methods), (c) ACE-PBE, and (d) ACE-\SCAN.
    For comparison, (a) and (b) also show the CALPHAD assessment as dotted lines by \cite{Sundman1998}\cite{Otis2024} using \texttt{pycalphad}\cite{Otis2017}.
    The statistical uncertainty of the free energies corresponds to an uncertainty of the phase boundaries of at most 10--20~K (Supplementary Note~4).
    }
    \label{fig:pd1}
\end{figure*}

We now compare the complete phase diagrams obtained from the atomistic free energy calculations. For each functional, we consider the phases identified on the corresponding convex hull together with the liquid phase. The temperature and composition ranges for the individual phases are chosen manually, while the remaining workflow is automated within \texttt{calphy}. Based on these inputs, \texttt{calphy} generates the required temperature-composition grid and performs the free energy calculations for each phase at the relevant state points. For stoichiometric phases, direct nonequilibrium Hamiltonian interpolation is employed, whereas for off-stoichiometric phases the alchemical transformations described in the Methods section are used.

For all phases, we use a temperature grid spacing of 25 K. For the FCC solid solution and the liquid phase, the composition interval is 0.02, while for \AuFourCuTwo, AuCu, and \AuCuThree~a finer composition interval of 0.01 is employed. The calculations are carried out in a high-throughput manner. At thermodynamic conditions where a phase is no longer metastable, the corresponding calculations may fail and are discarded. The resulting free energies are then passed to \texttt{landau}, which performs the semi-grandcanonical analysis described in the Methods and constructs the phase diagram.

We first consider the phase diagram obtained with the ACE-LDA potential, shown in Fig.~\ref{fig:pd1}(a). The calculated phase diagram reproduces the main topological features of the experimental and CALPHAD-assessed AuCu phase diagram \cite{Sundman1998}: an FCC solid solution at high temperature, three ordered intermetallics on the Au-rich, equiatomic and Cu-rich sides, and a continuous liquidus across the composition range. 
The quantitative agreement varies across the diagram.
The order-disorder transition lies about 50~K above the CALPHAD assessment, whereas the Cu melting temperature and the liquidus at high Cu content are both overestimated by about 110~K.
The other two functionals describe the Cu melting point and the nearby liquidus better, but ACE-LDA describes the liquidus better toward the Au-rich side of the phase diagram.
Despite these deviations, the present framework recovers the main features of the AuCu phase diagram directly from atomistic free energies, without empirical fitting of thermodynamic models.
Perhaps the most notable difference from the experimental and CALPHAD phase diagrams is the absence of the Au$_3$Cu phase. None of the DFT functionals considered here predict Au$_3$Cu to be stable on the convex hull; instead, \AuFourCuTwo~is found as the stable ordered compound on the Au-rich side. We also performed explicit free energy calculations for Au$_3$Cu and confirmed that it does not become stable at finite temperature within the present description, suggesting that the difference originates already at the level of the underlying energetics rather than from the finite-temperature treatment. We also note that CALPHAD descriptions have historically not included the \AuFourCuTwo~phase.

In Fig.~\ref{fig:pd1}(b), we show the corresponding phase diagram obtained with the same ACE-LDA potential but without explicit configurational sampling (the no-swap calculation; see Methods). Several significant differences are observed. The most important change is the much narrower stability range in temperature of the FCC solid solution. The ideal approximation clearly underestimates this range. This reflects not only the absence of explicit configurational entropy beyond ideal mixing, but also the fact that the energetic contribution associated with local chemical rearrangements is sampled more accurately when identity exchange moves are included. Because the non-ideal mixing entropy can only ever be smaller than the ideal one, it follows directly that there are significant, energetically favorable, enthalpic contributions in the solid solution. Furthermore, the solubility ranges of \AuFourCuTwo, AuCu, and \AuCuThree~are also strongly affected. Without explicit configurational sampling, these phases appear much closer to line compounds, whereas the calculations with identity exchange moves recover finite solubility ranges. Explicit configurational sampling is, therefore, essential not only for locating transition temperatures, but also for describing the extent of phase fields in composition space. 

We finally examine the exchange-correlation functional dependence, which constitutes a central limitation on the quantitative predictive accuracy of the phase diagrams. The phase diagrams obtained with ACE-PBE and ACE-\SCAN{} are shown in Fig.~\ref{fig:pd1}(c) and Fig.~\ref{fig:pd1}(d), respectively. The differences between the functionals are reflected most clearly in the melting temperatures, as discussed in the previous section, and therefore also in the liquidus and solidus lines. Differences are also observed in the order-disorder transition temperatures and in the detailed topology of the phase boundaries. The comparison therefore shows that, in addition to configurational entropy, the choice of the underlying DFT functional has a pronounced impact on the predicted finite-temperature thermodynamics. 
The differences between the three phase diagrams are substantially larger than the propagated uncertainty of the free energy calculations (Supplementary Note~4) and than the fitting error of the potentials measured by free energy perturbation (Methods Sec.~\ref{sec:fep}).
This highlights the sensitivity of the predicted phase boundaries to the underlying exchange-correlation functional and shows that no single functional provides uniformly better agreement across all thermodynamic observables considered here.

Comparing the two ACE-LDA phase diagrams in Fig.~\ref{fig:pd1}(a) and (b),
only ideal mixing approximation substantially underestimates the stability of the FCC solid solution against both the liquid and the ordered intermetallics, while both variants overestimate the depth of the liquidus minimum at intermediate compositions relative to the CALPHAD assessment.
As discussed above, explicit configurational sampling markedly improves the shape of the predicted binodals. 
Apart from the \AuFourCuTwo~phase, which is absent from the CALPHAD description, and the liquidus and solidus on the Cu-rich side, the ACE-LDA binodals follow the topology of the CALPHAD assessment with an approximately systematic temperature offset.
Finally, we note that an even more complete description of solubility ranges of the intermetallics may require the explicit inclusion of additional defects, such as vacancies, which we do not include in our framework.

\section{Discussion}

In this work, we have developed an atomistic free energy framework for multicomponent materials that explicitly incorporates configurational entropy within thermodynamic integration. By combining nonequilibrium alchemical transformations with atomic identity exchange moves, the method enables vibrational and configurational contributions to be treated within a single calculation. 

The methodology has been implemented in an automated workflow \cite{pyiron-paper, Menon2024} in which the free energy calculations, and the construction of phase diagrams can be carried out in a high-throughput manner. Starting from a given interatomic potential and a set of candidate phases, the workflow automatically evaluates free energies over the relevant temperature-composition range and constructs the corresponding phase diagram. 

Using AuCu as a representative example, we have applied the method to investigate order-disorder thermodynamics, melting transition, and the full composition-temperature phase diagram. Explicit configurational sampling lowers the AuCu order-disorder transition relative to the ideal-mixing approximation. In particular, configurational sampling affects not only the entropic stabilization of disordered phases, but also the effective free energy balance between competing ordered and disordered states through the sampling of chemically distinct local environments. Approaches based only on ideal mixing can therefore miss important features of the phase diagram, especially in systems with strong ordering tendencies or non-ideal chemical correlations.

Our methodology also enables the sensitivity of finite temperature thermodynamic predictions to the underlying DFT description to be assessed systematically. By training ACE potentials on data generated with different exchange-correlation functionals, we were able to compare how the choice of functional influences melting temperatures, order-disorder transitions, and the resulting phase boundaries. Such a direct comparison at the level of full phase diagrams would be prohibitively expensive with DFT alone. A clear outcome of this comparison is that no single functional is uniformly best: LDA describes the Au melting temperature most accurately as compared to experiments, \SCAN{} the Cu melting temperature, and the LDA Cu-rich liquidus is overestimated by $\sim$110~K owing to the overestimated Cu melting point. This non-uniformity cautions against selecting a functional on the basis of a single property and motivates assessing finite-temperature thermodynamics across functionals.
Therefore, for a new material system, the choice of exchange-correlation functional should first reflect the relevant chemistry and bonding, together with the known strengths and limitations of the available approximations for that class of materials.
The choice could then be benchmarked against quantities that control the competing phases, such as relative phase energies, mixing energetics, structural properties, and, where available, experimental thermodynamic data including melting or ordering temperatures.
With the present workflow, the melting temperatures of the elements are computationally efficient to obtain and are usually known experimentally, so they provide a natural first test of a functional, and free energy perturbation separates the error of the interatomic potential from that of the exchange-correlation functional.

Furthermore, the present approach provides a way to complement existing CALPHAD assessments and experimental phase-diagram data with high-accuracy atomistic predictions based on machine-learning interatomic potentials. Rather than replacing these established approaches, it offers a direct route from the underlying atomic interactions to finite-temperature phase stability, and is therefore particularly useful in systems where experimental information is sparse or where thermodynamic descriptions remain uncertain. In this sense, the framework provides a foundation for predictive phase discovery and accelerated materials design, especially when combined with high throughput workflows and universal machine learning interatomic potentials.

\section{Methods}

\begin{figure*}[htp!]
    \centering
    \includegraphics[width=0.8\linewidth]{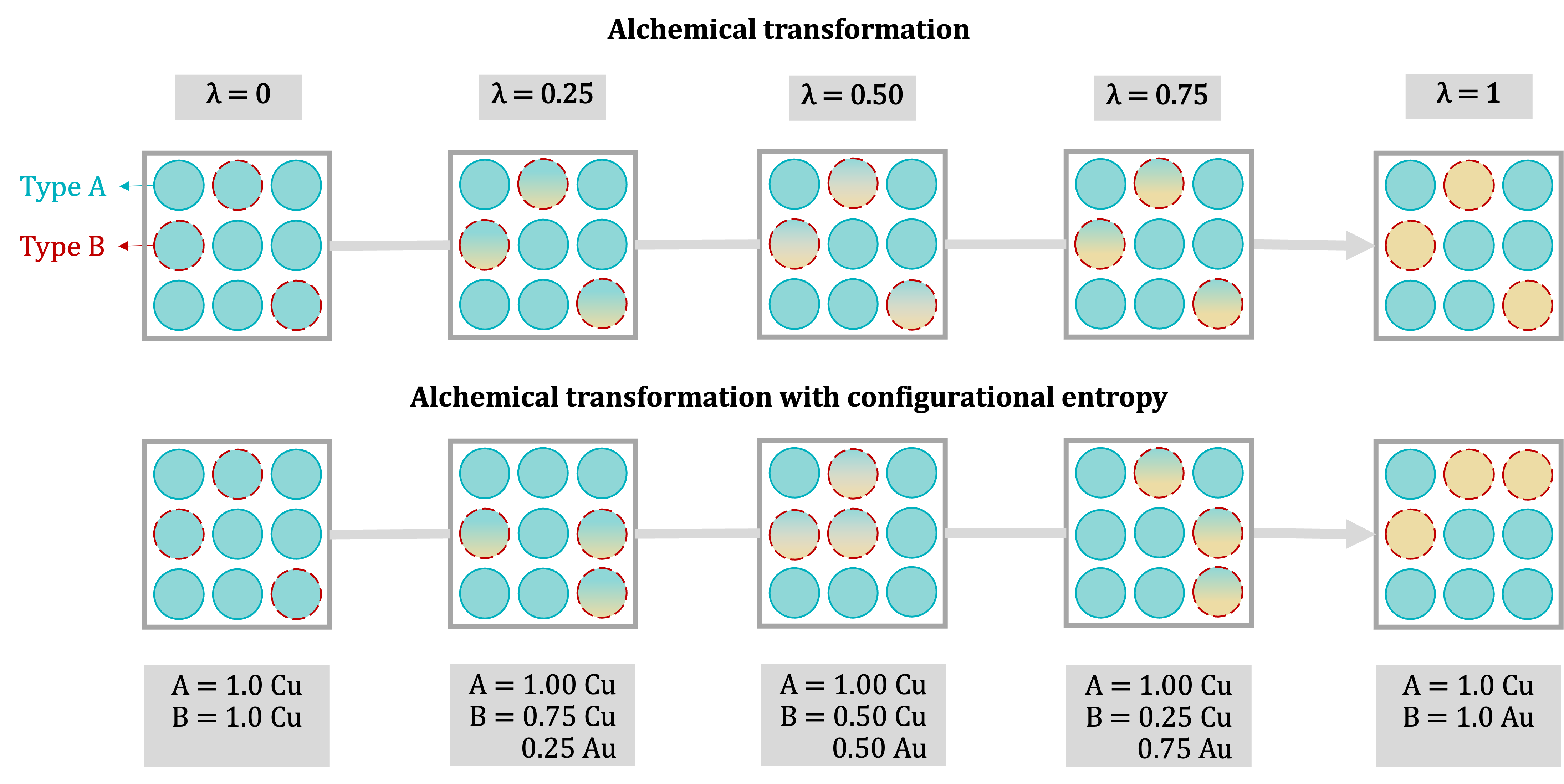}
    \caption{
    Schematic of the alchemical transformation path on a fixed set of sites. Sites designated as type $B$ (red outline) have their interactions continuously interpolated from $A$ (Cu, cyan, $\lambda=0$) to $B$ (Au, orange, $\lambda=1$). (Top) Interpolation at fixed site assignment. This approach, along with ideal mixing approximation, is used for the calculation of phase diagrams which do not include configurational entropy. (Bottom) The same interpolation combined with Monte Carlo identity-exchange moves, which reassign the $B$ label among sites between MD blocks to sample the configurational distribution.
    }
    \label{fig:overviewmethod}
\end{figure*}

\subsection{Non-equilibrium thermodynamic integration for binary alloys} \label{sec:thermo}

We show the non-equilibrium thermodynamic integration formalism from Ref.~\cite{Menon2021} and specialize it to a binary system in which the composition is varied along the integration path at fixed temperature.

We consider two Hamiltonians $H_i$ and $H_f$ connected by a linear interpolation,
\begin{equation}H(\lambda) = (1-\lambda) H_i + \lambda H_f ,
\label{eq:Hpath-new}\end{equation}
with interpolation parameter $\lambda \in [0,1]$. For a switching trajectory of duration $t_{\mathrm{sw}}$, $\lambda(t=0) = 0$ and $\lambda(t=t_{\mathrm{sw}}) = 1$, the dynamic work done by varying $\lambda$ in time is
\begin{equation}W^{s} = \int_{0}^{t_{\mathrm{sw}}} \frac{d \lambda}{dt} \, \frac{\partial H(\lambda)}{\partial \lambda} \, dt .
\label{eq:Ws-new}\end{equation}
Averaging forward and reverse switching trajectories removes the leading dissipative contribution, and the free energy difference between the two endpoints is recovered as
\begin{equation}\Delta F = F_f - F_i = \tfrac{1}{2} \left[ W^{s}_{i \to f} - W^{s}_{f \to i} \right] .
\label{eq:dFsym-new}\end{equation}
Equations~(\ref{eq:Hpath-new})-(\ref{eq:dFsym-new}) form the basis of the algorithms implemented in \texttt{calphy}~\cite{Menon2021}.

For a binary alloy of species $A$ and $B$ at fixed temperature $T$ and total atom number $N=N_A+N_B$, with composition $x = N_B/N$ and atomic masses $m_A$ and $m_B$. The Helmholtz free energy can be written as
\begin{align}F(T,x)
&= F_{\mathrm{int}}(T,x) \nonumber \\
&\quad + k_B T N \left[ (1-x) \ln (1-x) + x \ln x \right] \nonumber \\
&\quad + 3 k_B T N \left[ (1-x) \ln \Lambda_A + x \ln \Lambda_B \right] \nonumber \\
&\quad + k_B T N (\ln N - 1) ,
\label{eq:Fbinary-new}\end{align}
where $\Lambda_A$ and $\Lambda_B$ are the thermal de Broglie wavelengths of species $A$ and $B$, given by $\Lambda_A = \sqrt{h^2/(2 \pi m_A k_B T)}$ and analogously for $\Lambda_B$. The second line is the ideal-mixing contribution arising from the indistinguishability prefactor $1/(N_A! N_B!)$ in the partition function and is independent of the interatomic potential. The third line is the kinetic-energy contribution evaluated with the species-resolved masses. The fourth line, $k_B T N(\ln N - 1)$, is the Stirling approximation of the indistinguishability factor $1/N!$ for the full system.

We consider an integration path that connects two binary states at compositions $x_i$ and $x_f$, with temperature kept fixed at $T$ throughout. The reference composition $x_i$ need not correspond to a pure element; the construction applies equally to a pure-$A$ reference ($x_i = 0$), a pure-$B$ reference ($x_i = 1$), or any intermediate composition $x_i \in (0,1)$ for which $F(T, x_i)$ is already known. Combining Eq.~(\ref{eq:Hpath-new}) with Eq.~(\ref{eq:Fbinary-new}), the free energy of the target state is expressed as
\begin{align}F(T, x_f)
&= F(T, x_i) \nonumber \\
&\quad + k_B T N \left[ (1-x_f) \ln (1-x_f) + x_f \ln x_f \right. \nonumber \\
& \qquad \qquad \quad \left. - (1-x_i) \ln (1-x_i) - x_i \ln x_i \right] \nonumber \\
&\quad + 3 k_B T N (x_f - x_i) \left[ \ln \Lambda_B - \ln \Lambda_A \right] \nonumber \\
&\quad + \int_{0}^{t_{\mathrm{sw}}} \frac{d \lambda}{dt} \, \frac{\partial U(\lambda)}{\partial \lambda} \, dt ,
\label{eq:Fbintarget-new}\end{align}
where the analytic mixing and de Broglie terms are evaluated directly from $x_i$, $x_f$ and the species masses, and the work integral is the binary specialization of Eq.~(\ref{eq:Ws-new}) along the switching trajectory connecting $H_i$ and $H_f$. The interaction-dependent term must include sampling of both the atomic positions $\pmb{r}^N$ and the assignment of species identities $\pmb{\sigma}^N$ to lattice sites. Standard molecular dynamics samples only $\pmb{r}^N$ at fixed $\pmb{\sigma}^N$; the extension required to sample $\pmb{\sigma}^N$ is described in the following section.

The simplest and most common choice is $x_i = 0$, since the free energy of a pure element is readily obtained from the standard Frenkel-Ladd path~\cite{Menon2021}. Once $F(T, x_f)$ has been computed for one composition, it can in turn serve as the reference for a subsequent transformation, enabling a chain of calculations across the composition range that need not return to a pure-element starting point.

We restrict the discussion to a binary system for clarity. Generalization to $k>2$ species is straightforward: the mixing term becomes a multinomial sum $\sum_i c_i \ln c_i$, the de Broglie contribution is summed over all species, and the identity-exchange moves introduced in the next subsection are extended to pick two atoms of any two distinct species.

\subsection{Composition path with identity-exchange sampling} \label{sec:conf}

We specify $H_i$ and $H_f$ for the binary transformation and describe how the configurational degrees of freedom are sampled along the integration path. We first describe the construction for $x_i = 0$, that is, a transformation from a pure-$A$ reference state to a binary target at composition $x_f = x$, $A \to A_{1-x} B_x$, at fixed temperature $T$. The extension to an arbitrary reference composition is summarized at the end of the section.

\subsubsection{Endpoint Hamiltonians}
Both endpoints are simulated on the same set of $N$ atomic sites and with the same assignment of identities matching the target composition: $N_A = (1-x) N$ atoms are assigned identity $A$ and $N_B = x N$ atoms are assigned identity $B$, distributed randomly over the lattice.
During the alchemical transformation, all atoms carry the same reference mass $m_{\mathrm{ref}}$, independent of their species identity.
Because the momentum contribution to the classical partition function factorizes from the configurational one, the physical species masses enter only through the analytic de Broglie term in Eq.~(\ref{eq:Fbintarget-new}), which restores them exactly.
The two endpoint Hamiltonians differ only in how the identities enter the potential:
\begin{itemize}
\item $H_i$: all interactions are evaluated with the pure-$A$ potential, $U_i(\pmb{r}^N) = U^{(A)}(\pmb{r}^N)$, so that the $B$ atoms are present in name only and do not feel any chemistry-specific interaction.
\item $H_f$: interactions are evaluated with the full binary potential and the assigned identities, $U_f(\pmb{r}^N, \pmb{\sigma}^N) = U^{(AB)}(\pmb{r}^N, \pmb{\sigma}^N)$.
\end{itemize}
Along the integration path the interpolated potential is
\begin{equation}U(\lambda) = (1-\lambda) U_i + \lambda U_f .
\label{eq:Ulambda-new}\end{equation}
Physically, the $N_B$ atoms designated as species $B$ begin (at $\lambda=0$) interacting exactly as $A$ atoms and are gradually transformed into $B$ atoms as $\lambda \to 1$. The path thus slowly converts $N_B$ of the $A$ atoms into $B$, as shown in Fig. \ref{fig:overviewmethod}. This continuous transformation of the $B$ identity is implemented in \textsc{lammps} with \texttt{pair\_style hybrid/scaled}, which weights the two endpoint potentials by $1-\lambda$ and $\lambda$.

For a transformation between two arbitrary binary compositions $x_i$ and $x_f$, the path slowly converts $(x_f - x_i)N$ atoms from $A$ into $B$ (for $x_f > x_i$), leaving the remaining atoms unchanged. 
Therefore, the same construction applies with a minor adjustment that the identities are assigned according to the target composition $x_f$, while all atoms retain the common reference mass $m_{\mathrm{ref}}$ throughout the switching calculation.
At $\lambda=0$, the potential treats only a subset of the atoms as $B$, corresponding to the reference composition $x_i$.
Once again, the physical mass dependence between the reference and target compositions is included analytically through the de Broglie correction in Eq.~(\ref{eq:Fbintarget-new}), and all subsequent sampling considerations carry over unchanged.

\subsubsection{Configurational sampling}

At each $\lambda \in (0,1)$, the canonical ensemble defined by $U(\lambda)$ involves both atomic positions $\pmb{r}^N$ and identity assignments $\pmb{\sigma}^N$ at fixed $(N_A, N_B)$. To sample $\pmb{\sigma}^N$ within the non-equilibrium switching protocol, we perform atomic identity-exchange moves between molecular-dynamics segments, as shown in bottom panel of Fig. \ref{fig:overviewmethod}. 
A trial swap is constructed by picking one atom of type $A$ and one of type $B$ at random and exchanging their species identities at fixed positions and velocities.
Because all atoms carry the same reference mass, the momenta and kinetic energies are also unchanged by the identity exchange.
The Metropolis acceptance criterion therefore depends only on the change in potential energy.
The swap is accepted or rejected according to the standard Metropolis criterion~\cite{Metropolis1953,Sadigh2012}.
In LAMMPS, the swaps are performed with \texttt{fix atom/swap} without kinetic energy rescaling (\texttt{ke no}).

\subsubsection{Combined MD/MC switching protocol}

The combined molecular dynamics / Monte Carlo non-equilibrium switching protocol is summarized in Algorithm~\ref{algo:swap-alch} and is controlled by three parameters: the total number of identity-exchange attempts per switching trajectory $n_{\mathrm{swap}}$, the switching time $t_{\mathrm{sw}}$, and the number of MD steps $n_{\mathrm{MD}}$ between successive swap blocks. The convergence of the protocol with respect to these parameters is reported in the Results section.

\begin{algorithm}[H]
\caption{Alchemical sampling with identity-exchange moves}
\label{algo:swap-alch}
\begin{algorithmic}[1]
\State assign identities: $N_A = (1-x) N$ as $A$, $N_B = x N$ as $B$
\State set up $H_i$ (pure-$A$ potential) and $H_f$ (binary $A$-$B$ potential)
\For{$n$ independent runs}
  \State equilibrate at $\lambda = 0$ for time $t_{\mathrm{eq}}$
  \State $n_{\mathrm{block}} = \lceil t_{\mathrm{sw}} / (n_{\mathrm{MD}} \Delta t) \rceil$
  \State $n_{\mathrm{sw/block}} = n_{\mathrm{swap}} / n_{\mathrm{block}}$
  \For{$k = 1, \ldots, n_{\mathrm{block}}$}
    \State run $n_{\mathrm{MD}}$ MD steps with $U(\lambda)$, ramping $\lambda$ continuously at rate $\dot\lambda = 1/t_{\mathrm{sw}}$
    \State accumulate $W^{s}_{i \to f}$ from Eq.~(\ref{eq:Ws-new})
    \State attempt $n_{\mathrm{sw/block}}$ identity-exchange moves at the current $\lambda$; accept or reject by the Metropolis criterion
  \EndFor
  \State equilibrate at $\lambda = 1$ for time $t_{\mathrm{eq}}$
  \State repeat the loop in reverse to accumulate $W^{s}_{f \to i}$
\EndFor
\State $\Delta F = \tfrac{1}{2} \langle W^{s}_{i \to f} - W^{s}_{f \to i} \rangle$
\State evaluate $F(T,x)$ from Eq.~(\ref{eq:Fbintarget-new}) using the analytic mixing and de Broglie terms
\end{algorithmic}
\end{algorithm}

Starting from a reference state at composition $x_i$ with known free energy~\cite{Menon2021}, Algorithm~\ref{algo:swap-alch} provides the free energy of the target binary state at temperature $T$ and composition $x_f$ with both vibrational and configurational degrees of freedom sampled within a single calculation.

\subsection{Phase diagram construction from atomistic free energies} \label{sec:landau}

Given the Helmholtz free energy per atom $f(c,T)$ of competing phases at fixed composition $c$ and temperature $T$, phase stability can be determined by transforming to the semi-grand potential. For a binary alloy, this potential is defined as
\begin{equation}\phi(\Delta\mu, T)
=
\min_{c}
\left\{
f(c,T) - c\,\Delta\mu
\right\},
\label{eq:sgc}\end{equation}
where $\Delta\mu = \mu_{\mathrm{Au}} - \mu_{\mathrm{Cu}}$ is the chemical potential difference between the two species. The stable phase at given $(\Delta\mu, T)$ is the one with the lowest value of $\phi$, the semi-grand potential.

Equation~(\ref{eq:sgc}) corresponds to a Legendre transformation of the Helmholtz free energy with respect to composition. This formulation is equivalent to the common tangent construction in $(c,T)$ space, but allows for a direct and numerically stable determination of phase stability on a grid of $(\Delta\mu, T)$ values.

For a set of candidate phases $\{\phi_i\}$, we independently evaluate the semi-grand potential for each phase. Phase boundaries are determined by identifying values of $(\Delta\mu, T)$ at which two phases yield equal $\phi$, corresponding to first-order transitions. Numerical root searches are performed to refine the transition lines.

The corresponding coexistence compositions in $(c,T)$ space are obtained from the thermodynamic identity
\begin{equation}c(\Delta\mu, T)
=
- \frac{\partial \phi(\Delta\mu, T)}{\partial \Delta\mu}.
\label{eq:c_from_phi}\end{equation}
The resulting coexistence points are mapped back to composition-temperature space to construct the phase diagram. This procedure enables a consistent construction of binary phase diagrams directly from atomistically computed free energies.

\subsection{Workflows for phase diagram construction} \label{sec:workflows}

The complete procedure for constructing phase diagrams from atomistic free energies is automated in a high-throughput workflow. For a binary alloy system, the required user inputs are limited to: (i) the interatomic potential, (ii) the set of candidate phases to be considered, and (iii) the temperature and composition ranges for each phase over which the free energies are to be calculated. All subsequent steps are performed automatically.

For each phase and composition, the workflow proceeds as follows. First, equilibrium structural properties are determined in the isothermal-isobaric ensemble. 
The simulation cells contain 4000 atoms for the FCC solid solution, the liquid and AuCu$_3$ ($10\times10\times10$ conventional FCC cells), and 3456 atoms for L1$_0$ AuCu and Au$_4$Cu$_2$.
All simulations use a timestep of 1~fs.
The equilibrium cell is obtained from 10~ps of simulation in the $NPT$ ensemble at zero pressure and the target temperature, using the Nos\'{e}--Hoover thermostat and barostat with damping times of 0.1~ps.
The cell shape is relaxed in a way that preserves the symmetry of each phase.
For the cubic phases and the liquid the three cell vectors are scaled together.
For the two tetragonal compounds, L1$_0$ AuCu and Au$_4$Cu$_2$, the in-plane lattice parameters remain equal, $a = b$, and the axial ratio $c/a$ relaxes freely, so that it takes its equilibrium value at every temperature and composition.
For solid phases, the mean-squared displacements $\langle (\Delta r)^2 \rangle$ are evaluated to determine the harmonic reference parameters required for the Einstein crystal construction. For liquid phases, the equilibrium density is determined and a suitable reference Hamiltonian is defined accordingly.

Using these quantities, a reference Hamiltonian $H_i$ is constructed and the corresponding reference free energy is evaluated. Thermodynamic integration is then performed to obtain the free energy of the interacting system at the desired temperature and composition, following the formalism described in Secs.~\ref{sec:thermo} and \ref{sec:conf}. 
The Frenkel--Ladd and liquid reference calculations switch over 25~ps in each direction at the averaged cell, using a Langevin thermostat with a damping time of 0.1~ps.
The alchemical composition transformations switch over 25~ps in each direction in the $NPT$ ensemble at zero pressure with the same Nos\'{e}--Hoover settings, after 10~ps of equilibration at each endpoint.
During the alchemical switching, both endpoint potentials are evaluated at every MD step.
The FCC solid solution and the liquid are sampled over the full composition range.
The ordered phases are sampled over $0.45 \le x_\mathrm{Cu} \le 0.55$ for L1$_0$ AuCu, $0.70 \le x_\mathrm{Cu} \le 0.80$ for AuCu$_3$, and $0.20 \le x_\mathrm{Cu} \le 0.40$ for Au$_4$Cu$_2$, in steps of about 0.01.

The free energy calculations are implemented in the \texttt{calphy} package \cite{Menon2021}, which provides automated reference construction, thermodynamic integration, and statistical averaging.

The resulting free energies $F(T,x)$ for all candidate phases are subsequently processed to determine phase stability using the semi-grandcanonical formulation described in Sec.~\ref{sec:landau}. The numerical evaluation of the Legendre transformation, identification of phase boundaries, and mapping of coexistence compositions are performed using the \texttt{landau} package \cite{Poul2025}, which implements the semi-grandcanonical minimization and root-search procedures required for constructing binary phase diagrams.

The separation between \texttt{calphy} and \texttt{landau} enables a modular workflow in which free energy computation and phase stability analysis can be performed independently. This structure facilitates extension to additional alloy systems and candidate phases without modification of the underlying formalism.

\subsection{ACE potentials for the AuCu system}

Interatomic interactions are modeled using a machine-learning interatomic potential constructed within the Atomic Cluster Expansion (ACE) framework \cite{Drautz2019, Lysogorskiy2021}. ACE expresses the total energy as a linear expansion in symmetry-adapted polynomial basis functions of local atomic environments, providing a systematically improvable and transferable representation of many-body interactions. When trained on density functional theory data, ACE potentials retain ab initio accuracy while enabling molecular dynamics simulations over the time and length scales required for free energy calculations.

\subsubsection{Training data generation with ASSYST\label{sec:assyst}}

Training data is generated with the ASSYST method as described in Ref.~\cite{Poul2025}.
It samples training structures from all possible space groups and applies multiple steps of relaxations before randomly sampling structures in the vicinity around these minima.
Notably, the structures contain very few atoms, which allows us to relabel structures with energies and forces from different functionals very efficiently. The entire dataset consists of around 32000 structures with 5 atoms per cell on average. The complete list of ASSYST parameters used for structure generation is given in Table \ref{tab:assyst_parameters}.
Generated structures with minimum atomic distance of less than 1.8 $\text{\r{A}}$ were filtered to avoid strong core repulsion.
The potentials are trained on all generated structures and validated on additional properties in the Supplementary Material. 

All structures were labelled with VASP\cite{Kresse1996a}\cite{Kresse1996b} (version 6.4) using the Projector Augmented Wave method\cite{Bloechl_1994}\cite{Kresse_1999}.
Both elements used the standard pseudopotentials from the VASP distribution, for the \SCAN{} calculations we used the same set as for the PBE calculations.
This corresponds to 3d$^{10}$ 4s$^1$-valent Cu and 5d$^{10}$ 6s$^1$-valent Au.
Except for the pseudopotentials and the electronic convergence criterion, all other parameters are kept fixed across the three functionals.
We used a plane wave cutoff of 550 eV with automatic $\Gamma$-centered $k$ meshes with minimum spacing of $0.1\cdot\text{\r{A}}^{-1}$ between points (\texttt{KSPACING=0.1}) and an electronic convergence criterion of $10^{-6}$ eV ($10^{-8}$ eV for PBE).
Occupations are smeared with the Methfessel-Paxton\cite{Methfessel_1989} scheme of order one with smearing parameter 0.2 eV.
The same settings were used for the verification calculations.
Datasets for all three functionals used the exact same structure set, except for rare, functional specific convergence problems or crashes.

\begin{table}
    \centering
    \caption{Input parameters used for generating the AuCu training dataset with the ASSYST workflow.}
    \label{tab:assyst_parameters}
    \begin{tabular}{ccl|c}
        \toprule
        \multicolumn{2}{c}{Step}   & parameter & value      \\
        \midrule
        Crystals& Au         & \#Atoms   & total $< 8$       \\
               &            & \#Samples & 841               \\
               & Cu         & \#Atoms   & total $< 8$       \\
               &            & \#Samples & 841               \\
               & Au/Cu      & \#Atoms/element  & 1,2,3,4,6         \\
               &            &           & total $< 8$       \\
               &            & \#Samples & 1335              \\
        \midrule
        \multicolumn{2}{c}{Relaxation}  & $n_\mathrm{trace}$        & 2        \\
        \midrule
        \multicolumn{2}{c}{Vibrational} & $\sigma_\mathrm{rattle}$  & 0.5\,\AA \\
                    &                   & $n_\mathrm{rattle}$       & 4        \\
        \midrule
        \multicolumn{2}{c}{Elastic} & $\epsilon_\mathrm{hydro}$ & 0.8 \\
                        &           & $\epsilon_\mathrm{shear}$ & 0.2 \\
                        &           & $n_\mathrm{stretch}$      & 2   \\
        \bottomrule
    \end{tabular}
\end{table}

\subsubsection{ACE hyperparameters}

ACE potentials were fitted with 1000 functions per element, an outer cutoff of 6.2~{\AA}, and a linear and Finnis-Sinclair output layer.
Structures were weighted by their distance from the training data convex hull.
The relative weight of forces to energies was set as 10\,\%.
We used L1-, and L2-norm regularization on the potential coefficients and all three available radial regularization schemes implemented in pacemaker \cite{Lysogorskiy2021}, all with regularization weights of $10^{-8}$.
Training proceeded according to the body-ordered ladder scheme, in steps of 50 basis functions or 5\,\% of the basis size, whichever is larger.
Each ladder step was run for a maximum of 1000 steps.
Core-repulsion is handled by switching smoothly to a ZBL potential in the range of 1.5-1.7~{\AA}.
The exact location of the cross-over point depends on each potential and was chosen to ensure continuously positive curvature of the pair interactions at close distances.

\subsubsection{Exchange-correlation functionals}

The choice of exchange-correlation functional is a key factor in the accuracy of DFT calculations. The Local Density Approximation (LDA) \cite{Perdew1981} and the Generalized Gradient Approximation (GGA) in the PBE parametrization \cite{Perdew1996} frequently show opposite trends for equilibrium properties of solids, with LDA overbinding and PBE underbinding, e.g.\ \cite{staroverov2004}\cite{Csonka2009} \cite{Zhang2018}. Similar trends are also reflected in thermodynamic properties such as melting temperatures \cite{Zhu2017}. In the case of Au and, in particular, the mixing enthalpies of AuCu, semi-local functionals are known to be especially challenging \cite{Zhang2014, Tian2016, Levamaki2018, Nepal2019}.
In Au, LDA has been shown\cite{Grabowski2015} to yield very good results, on par with much more expensive Random Phase Approximation (RPA) calculations. The more recent SCAN functional \cite{Sun2015}, and its regularized variant \SCAN{} \cite{Furness_2020}, have been shown to improve the description of lattice parameters and mixing enthalpies compared to earlier semi-local functionals. Here, we exploit the efficiency of the methods developed for potential fitting and free energy calculations to examine whether this improved description also persists at elevated temperatures and in the resulting phase boundaries, which would be prohibitively expensive to assess using direct DFT calculations alone.

For this purpose, we parameterize separate ACE potentials for each functional, denoted ACE-LDA, ACE-PBE, and ACE-\SCAN, using DFT datasets generated consistently within the corresponding exchange-correlation approximation. This enables a direct comparison of how the underlying functional influences the predicted thermodynamics through the resulting ACE potentials. The quality of the fits is assessed against the corresponding DFT reference data, and detailed benchmarks, including energy and force errors, are provided in the Supplementary Material.

\subsection{Free Energy Perturbation Calculations \label{sec:fep}}

Free Energy Perturbation (FEP)~\cite{zwanzig1954high,Grabowski2009} calculations were performed to establish the thermodynamic reliability of the ACE potentials and estimate the error in the presented phase diagrams.

Samples are drawn from equilibrium molecular dynamics with the ACE potential and reweighted with the energy difference between DFT and ACE, which gives the free energy difference between the two descriptions in each phase.
We run NPT MD near the transition temperatures (Table~\ref{tab:fepshift}) to establish the equilibrium cell parameters for a cell with 108--144 atoms. 
At these averaged cell parameters we perform NVT MD for 1 ns and take 16 uncorrelated samples, with the arrangement of the species kept fixed.
For each sample, we perform a single point DFT calculation with the same settings as in Sec. \ref{sec:assyst}, except we enable Fermi smearing with width $\sigma = k_B T$ to check on the influence of electronic temperature on the phase transitions.
From a set of samples the correction to the free energy at a given temperature can be calculated as

\begingroup
\begin{align}\Delta E_i &= E_{\mathrm{DFT},i} - E_{\mathrm{ACE},i} \nonumber \\
\Delta F &= \frac{\langle \Delta E \rangle}{N} - \frac{k_B T}{N} \log \left\langle
    \exp{ \left(-\frac{ \Delta E - \langle \Delta E \rangle }{k_B T}\right) }
    \right\rangle \label{eq:fep}\end{align}
\endgroup

where $i$ is the $i$th snapshot taken from the MD samples.

These calculations were performed for the melting points of pure Au and Cu with ACE-LDA, for the melting point of pure Au with ACE-PBE, and for L1$_0$ AuCu and the FCC solid solution at $x_\mathrm{Cu}=0.5$ with ACE-LDA near the order-disorder transition.
For the melting points, we added the corrections to the ACE free energies and recomputed the free energy crossings.

The results are given in Tabs.~\ref{tab:fepshift} and \ref{tab:feptemp}.
The absolute corrections for ACE-LDA are large because of a difference in the zero of energy between DFT and the training data of ACE-LDA.
For training ACE-LDA, all energies were shifted by 58\,meV per Au atom and 19\,meV per Cu atom, so that the energy of well separated atoms is zero.
This shift accounts almost entirely for the difference in magnitude between the LDA and PBE values in Tab.~\ref{tab:fepshift}.
The remaining differences between competing phases are 1.1--4.3\,meV/atom.
For melting, they raise the transition temperatures by 15--44\,K, which places the DFT melting points at 1183\,K (Au, LDA), 861\,K (Au, PBE) and 1512\,K (Cu, LDA).
These shifts are much smaller than the differences between the functionals.
For the order-disorder transition, DFT lowers the free energy of the FCC solid solution by 1.7\,meV/atom relative to L1$_0$ AuCu.
Since these calculations sample only the vibrations at a fixed arrangement of the species, this value cannot be interpreted as a shift of the transition temperature.
By itself, it would lower the transition temperature at the DFT level, but changes in the local ordering may shift the free energies of both phases in either direction.

\begin{table}
  \centering
  \caption{Corrections to ACE derived free energy as evaluated by Eq.~\ref{eq:fep} in meV/atom, with and without the electronic free energy $F_\mathrm{el}$ and the corresponding temperatures at which they were calculated.
  Values in parentheses give the error in the FEP estimator via the standard deviation of energy differences between snapshots.
  \label{tab:fepshift}
  }
  \begin{tabular}{llrrr}
        \toprule
        XC & Phase & $T$ & with $F_\mathrm{el}$ & no $F_\mathrm{el}$ \\
        & & [K] & \multicolumn{2}{c}{[meV/atom]} \\

        \midrule
        LDA & Au fcc & 1166 & $-$62.14(25) & $-$57.39(25) \\
        & Au liquid & 1166 & $-$60.70(56) & $-$55.80(57) \\
        & Cu fcc & 1467 & $-$27.89(28) & $-$20.10(29) \\
        & Cu liquid & 1467 & $-$23.54(41) & $-$15.47(46) \\
        & AuCu L1$_0$ & 721 & $-$37.68(23) & $-$35.81(23) \\
        & AuCu fcc & 721 & $-$39.36(39) & $-$37.51(39) \\
        \addlinespace
        PBE & Au fcc & 846 & $-$2.49(17) & 0.03(18) \\
        & Au liquid & 846 & $-$1.42(74) & 1.20(74) \\
        \bottomrule
  \end{tabular}

\end{table}

\begin{table}
  \centering
  \caption{Melting temperatures $T_m$ of the ACE potentials and their FEP correction $\Delta T$ to DFT, in K, with and without the electronic free energy $F_\mathrm{el}$.\label{tab:feptemp}}
  \begin{tabular}{llrrr}
    \toprule
       & & $T_m$ [K] & \multicolumn{2}{c}{$\Delta T$ [K]} \\ \cmidrule(lr){3-5}
      XC & Element & ACE & with $F_\mathrm{el}$ & no $F_\mathrm{el}$ \\
    \midrule
    LDA & Au & 1165.7 & +17.3 & +19.2 \\
        & Cu & 1467.3 & +44.3 & +47.1 \\
    \addlinespace
    PBE & Au & 845.5 & +15.0 & +16.5 \\
    \bottomrule
  \end{tabular}
\end{table}

On the same snapshots, the ACE forces agree with the DFT forces to within a few percent.

\section{Data availability}

The DFT training data and the resulting ACE potentials generated and analyzed during the current study are available from these repositories: \cite{Dataset2026} and \cite{Potentials2026}.
The VASP input files, the free energy perturbation data, the computed free energies and the scripts used to construct the phase diagrams and figures are available from \cite{Archive2026}.

\section{Code availability}

The free energy calculations in this work were carried out using the \href{https://github.com/ICAMS/calphy}{\texttt{calphy}} package.
Alchemical sampling including configurational entropy is implemented in release 2.0.0.
\texttt{calphy} used a customized version of LAMMPS \cite{Thompson2022} (version 11 Feb 2026), available at \url{https://github.com/thermoatoms/lammps}.
The \href{https://github.com/eisenforschung/landau}{\texttt{landau}} package is
used for phase-stability analysis and phase-diagram construction (version 1.16.0, \href{https://doi.org/10.5281/zenodo.23247372}{doi:10.5281/zenodo.23247372}). 
ACE interatomic potentials were fitted using \href{https://github.com/ICAMS/python-ace}{\texttt{pacemaker}} (version 0.3.0).

\bibliographystyle{simplejournal}
\bibliography{references}

\end{document}

\maketitle

\section{Local-energy evaluation of identity-exchange moves}
\label{si:local-energy}

The combined MD/MC switching protocol attempts a large number of identity-exchange moves along each transformation path, and the evaluation of the energy change associated with every trial swap dominates the overall cost. Here we describe how this cost is reduced from a scaling linear in the system size to one independent of it. The corresponding implementation is provided in a customized version of LAMMPS \cite{Thompson2022}, available at \url{https://github.com/thermoatoms/lammps}.

A trial identity-exchange move exchanges the species identities $\sigma_a$ and $\sigma_b$ of two atoms $a$ and $b$ at fixed positions and velocities (since all atoms carry the same mass during the transformation, see Methods Sec.~4.2, the kinetic energy is unchanged), and is accepted with the Metropolis probability
\begin{equation}P_{\mathrm{acc}} = \min\!\left[1,\, e^{-\beta \Delta E}\right],
\qquad
\Delta E = E(\pmb{\sigma}') - E(\pmb{\sigma}),
\label{eq:metropolis}\end{equation}
where $\pmb{\sigma}$ and $\pmb{\sigma}'$ denote the identity configurations before and after the swap and $\beta = 1/k_B T$. Only the energy difference $\Delta E$ enters the acceptance criterion; the forces are not required and are evaluated only on the molecular-dynamics integration steps. This avoids the full force evaluation that a standard MC step would otherwise trigger after each trial move, which for many-body machine-learning potentials such as ACE constitutes the dominant cost.

The key observation is that $\Delta E$ is a strictly local quantity. For a potential expressed as a sum of per-atom site energies,
\begin{equation}E(\pmb{\sigma}) = \sum_{i=1}^{N} \varepsilon_i\!\left(\{\pmb{r}_j, \sigma_j\}_{j \in \mathcal{N}_i}\right),
\label{eq:siteenergy}\end{equation}
where $\varepsilon_i$ depends only on the atoms in the neighbor shell $\mathcal{N}_i$ of atom $i$ within the cutoff $r_c$, a swap of the identities of atoms $a$ and $b$ changes $\varepsilon_i$ only for those atoms whose neighbor shell contains $a$ or $b$. Defining the affected set
\begin{equation}\mathcal{A} = \{a, b\} \cup \{\, i : a \in \mathcal{N}_i \ \text{or}\ b \in \mathcal{N}_i \,\},
\label{eq:affected}\end{equation}
the energy difference reduces to a sum over $\mathcal{A}$ alone,
\begin{equation}\Delta E = \sum_{i \in \mathcal{A}}
\left[ \varepsilon_i(\pmb{\sigma}') - \varepsilon_i(\pmb{\sigma}) \right].
\label{eq:deltaE_local}\end{equation}
The total energy is maintained as a cached quantity and updated incrementally by $\Delta E$ after each accepted move, so that no global energy evaluation is performed during the swap sequence. Since $|\mathcal{A}|$ is bounded by the local coordination and is independent of $N$, the cost per trial swap is reduced from $\mathcal{O}(N)$ to $\mathcal{O}(1)$. The set $\mathcal{A}$ is constructed from a full neighbor list, which identifies every atom $i$ that has $a$ or $b$ as a neighbor.

For potentials in which $\varepsilon_i$ is an explicit function of the local environment, such as ACE, Eq.~(\ref{eq:affected}) with the first coordination shell captures the change exactly. We verified that the local and full evaluations yield identical free energy differences to within statistical uncertainty, confirming that the optimization introduces no systematic error.

\section{Comparison with SGC-MC/MD}
\label{si:sgc}

\begin{figure*}[htp!]
    \centering
    \includegraphics[width=1.0\linewidth]{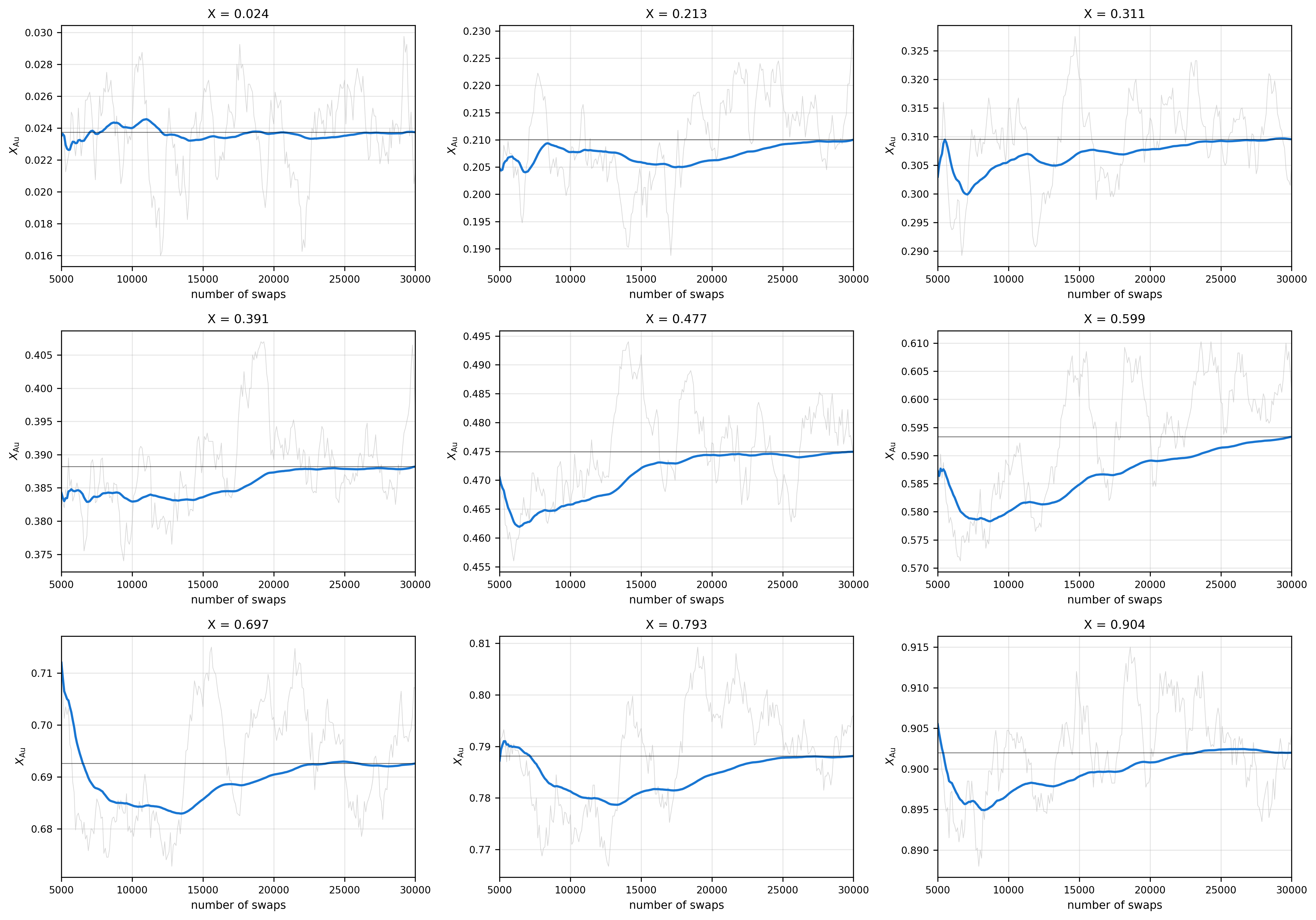}
    \caption{
    Convergence of the Au fraction $x_\mathrm{Au}$ in the SGC-MC/MD simulations of the AuCu FCC solid solution at 1000~K, shown as a function of the number of identity-exchange (swap) moves. Each panel corresponds to a different target composition, ranging from $x_\mathrm{Au}=0.024$ to $x_\mathrm{Au}=0.904$. The gray line shows the instantaneous value, the blue line the cumulative running average, and the horizontal line the converged mean obtained at the end of the run. 
    }
    \label{fig:sgc_x}
\end{figure*}

\begin{figure*}[htp!]
    \centering
    \includegraphics[width=1.0\linewidth]{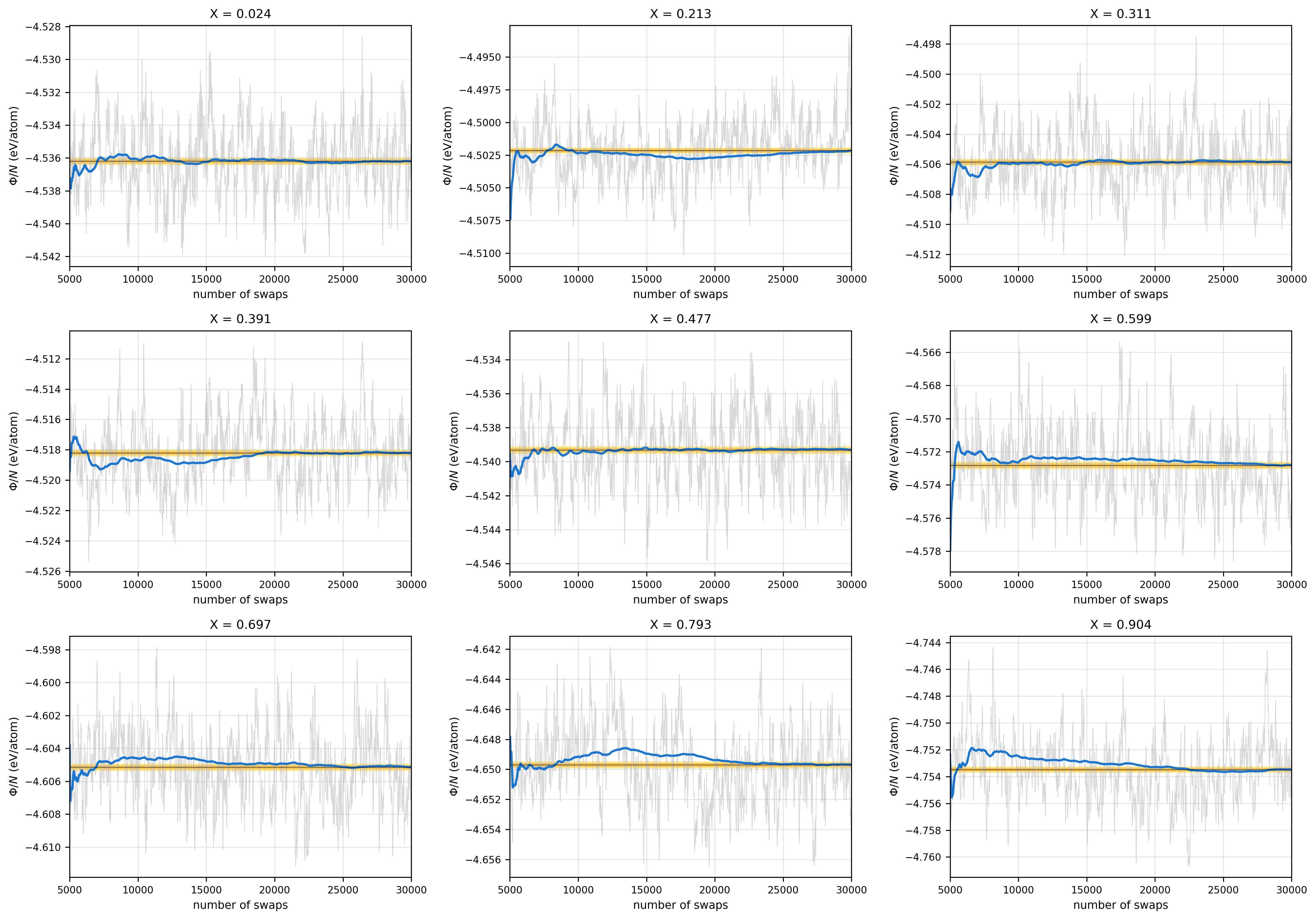}
    \caption{
    Convergence of the semi-grand canonical potential $\phi$ (in eV/atom) in the SGC-MC/MD simulations of the AuCu FCC solid solution at 1000~K, shown as a function of the number of identity-exchange (swap) moves, for the same compositions as in Fig.~\ref{fig:sgc_x}. The gray line shows the instantaneous value and the blue line the cumulative running average; the horizontal line indicates the converged mean obtained at the end of the run.
    }
    \label{fig:sgc_phi}
\end{figure*}

To benchmark the alchemical transformation approach, we compare it against semi-grand canonical Monte Carlo / molecular dynamics (SGC-MC/MD) simulations of the AuCu FCC solid solution at 1000~K.
We choose this temperature because the FCC solid solution is stable across the full composition range.
The SGC-MC/MD calculations use systems containing 4000 atoms and 30 values of the applied chemical potential difference $\Delta\mu$.
The convergence of the SGC-MC/MD sampling is shown in Figs.~\ref{fig:sgc_x} and \ref{fig:sgc_phi} for a representative set of compositions spanning the full range.
In Fig. \ref{fig:sgc_x} we show the convergence of the Au fraction $x_\mathrm{Au}$ in the SGC-MC/MD while in Fig. \ref{fig:sgc_phi} the corresponding semi-grand potential is shown. In all cases the running average settles onto the converged mean within the simulated number of swaps.
Because the composition depends exponentially on $\Delta\mu$ near the pure elements, the $\Delta\mu$ grid is refined at both ends, so that the integrated free energy does not depend on the integration scheme.
As a further check, six of the $\Delta\mu$ points were repeated starting from pure Cu instead of pure Au and reproduce the equilibrium compositions within 0.002, and seven points were continued for an additional 30\,000 swap steps without changes beyond the statistical uncertainty.

\begin{figure*}[htp!]
    \centering
    \includegraphics[width=1.0\linewidth]{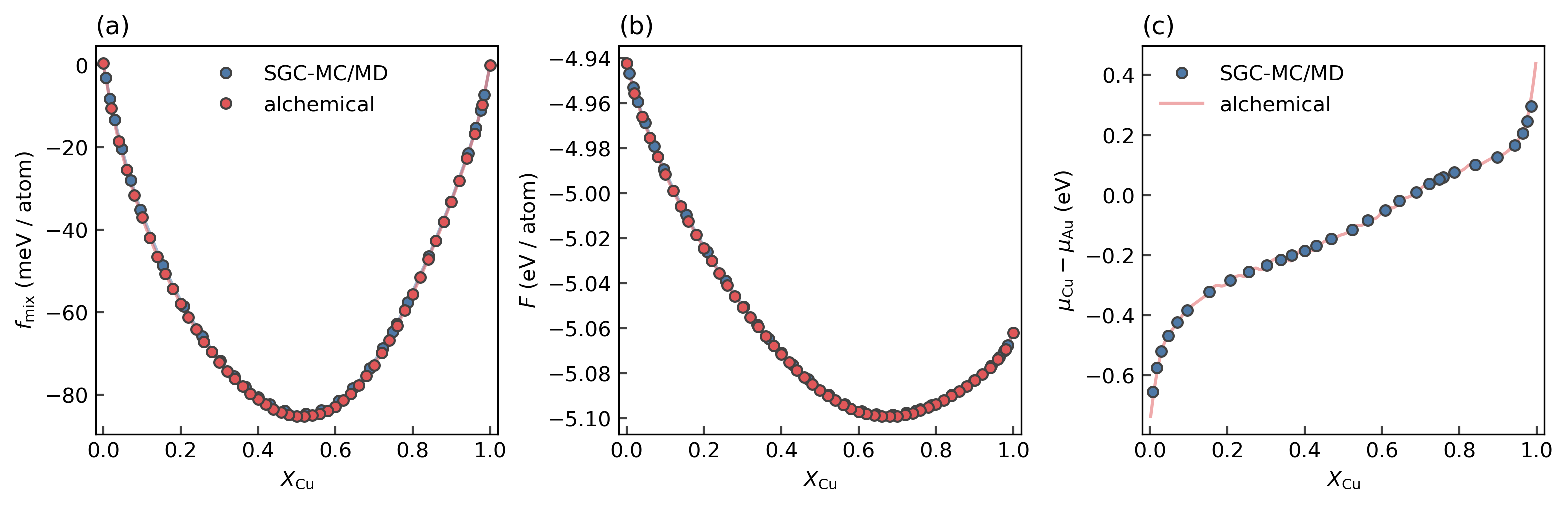}
    \caption{
    Comparison of alchemical transformations and SGC-MC/MD calculations for the AuCu FCC solid solution at 1000~K.
    (a) Free energy of mixing obtained with the two methods.
    (b) Absolute free energy, with the SGC-MC/MD results anchored to the directly calculated pure element free energies.
    (c) Chemical potential difference $\mu_{\mathrm{Cu}}-\mu_{\mathrm{Au}}$, obtained from the alchemical free energy and from the chemical potential imposed in the SGC-MC/MD calculations.
    }
    \label{fig:sgc_fe}
\end{figure*}

The free energy of mixing and the absolute free energy from the SGC-MC/MD calculations at 1000 K are compared with the alchemical transformations in Fig. \ref{fig:sgc_fe} (a) and (b), respectively.
For the comparison of absolute free energies, the SGC-MC/MD curve is anchored to the directly calculated pure element free energies used in the phase diagram construction.
The two methods agree on the free energy of mixing with an RMSE of 0.6~meV/atom over the full composition range and differ by $0.7\pm0.1$~meV/atom at $x_{\mathrm{Cu}}=0.5$.

We additionally compare the chemical potential difference obtained from the alchemical free energy as $-dF/dx_{\mathrm{Au}}$ with the value imposed in the corresponding SGC-MC/MD simulations in Fig. \ref{fig:sgc_fe} (c).
We add the kinetic contribution $\tfrac{3}{2}k_BT\ln(m_{\mathrm{Au}}/m_{\mathrm{Cu}})$ explicitly to the SGC-MC/MD chemical potentials, as this is already present in the alchemical calculations.
The chemical potentials agree within 6~meV.

\section{Dissipation in alchemical transformations}
\label{si:diss}

Each alchemical free energy difference is obtained from a forward (increase Au composition in Cu) and a reverse (decrease Au composition in Cu) switching run. Following the nonequilibrium thermodynamic integration formalism \cite{Menon2021}, the irreversible work accumulated along each direction, $W_{\mathrm{fwd}}$ and $W_{\mathrm{bwd}}$, combines into the reversible free energy difference
$$\Delta F = \tfrac{1}{2}\left(W_{\mathrm{fwd}} - W_{\mathrm{bwd}}\right),$$
while their sum measures the dissipated heat, i.e. the departure from a quasi-static path,
$$q = \tfrac{1}{2}\left(W_{\mathrm{fwd}} + W_{\mathrm{bwd}}\right).$$
Because the dissipation enters $W_{\mathrm{fwd}}$ and $W_{\mathrm{bwd}}$ with opposite sign in $\Delta F$, the bidirectional average cancels it to leading order.

\begin{figure}[htp!]
    \centering
    \includegraphics[width=1.0\linewidth]{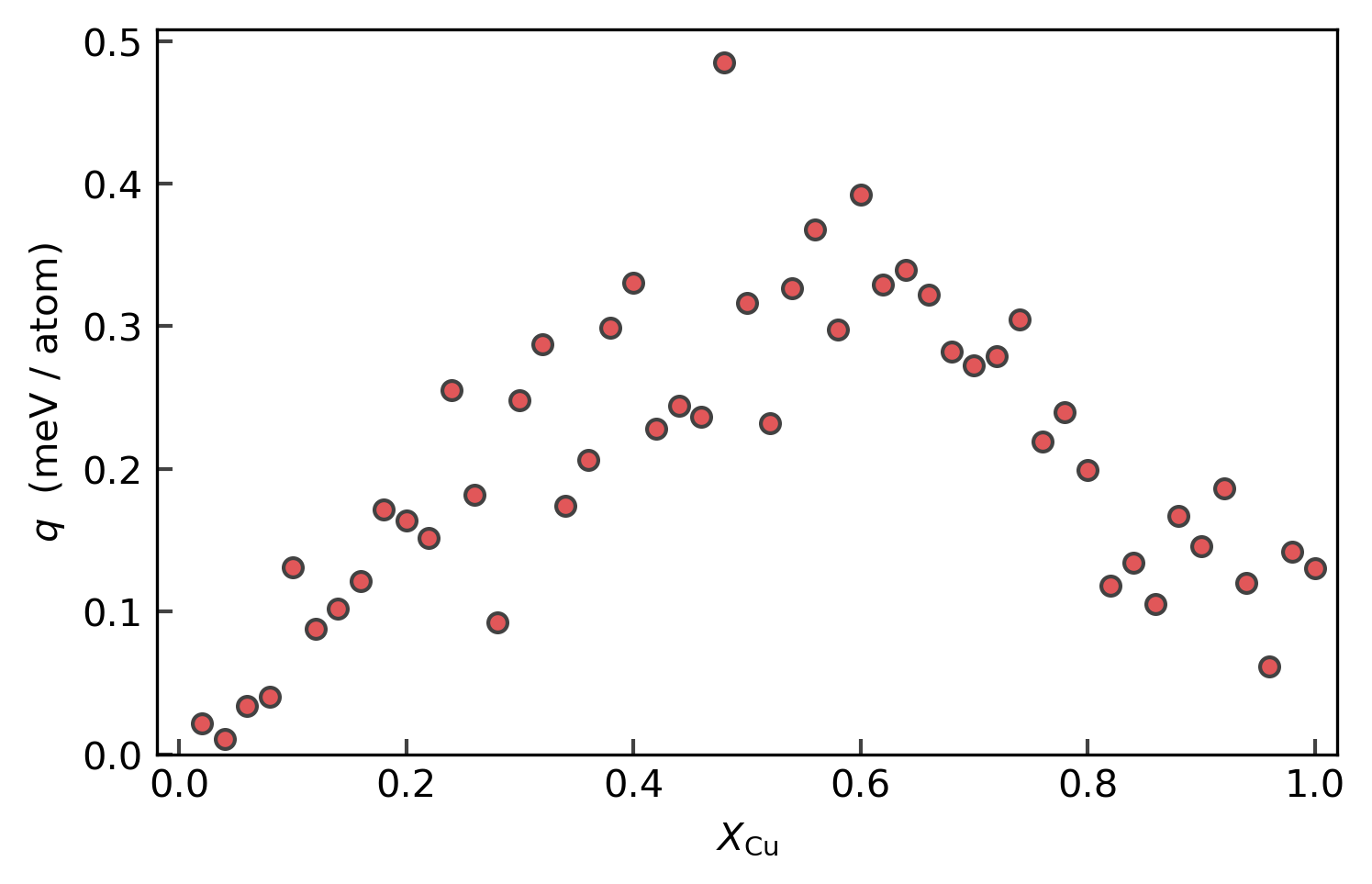}
    \caption{
    Per-point switching dissipation $q$ for the Cu–Au composition-scaling transformations at $1000~\mathrm{K}$, computed from the forward and reverse switching runs at each composition $x_{\mathrm{Au}}$. The dissipation stays below $0.5~\mathrm{meV/atom}$ throughout.
    }
    \label{fig:diss}
\end{figure}

Figure \ref{fig:diss} shows $q$ for the Cu–Au alchemical transformations at $1000~\mathrm{K}$. It remains below $0.5~\mathrm{meV/atom}$ at every composition.
Close to the order-disorder transition, at $x_\mathrm{Cu}=0.5$ and 750~K, the dissipation decreases from 2.3~meV/atom with 5000 swap attempts to 0.5~meV/atom at the 250\,000 attempts used in production, and remains at about 0.46~meV/atom up to 750\,000 attempts.
Over the same range the free energy difference converges (Fig.~\ref{fig:convergence_states}).
In the dilute solid solution at $x_\mathrm{Cu}=0.02$ the dissipation stays below 0.03~meV/atom for any number of swap attempts.
More frequent sampling of the arrangement of the species therefore reduces the dissipation, and the free energy from the bidirectional average agrees with the equilibrium SGC-MC/MD reference (Supplementary Note~2).

\section{Sensitivity of transition temperatures to free energy uncertainty}

To assess how uncertainties in the computed free energies propagate to the transition temperatures, we performed a first-order sensitivity analysis.
At the phase transition, the two competing phases satisfy $G_\alpha(T_0) = G_\beta(T_0)$, where $T_0$ is the transition temperature.
The free energy difference is $\Delta G(T) = G_\alpha(T) - G_\beta(T)$.
A small perturbation $\delta \Delta G$ shifts the crossing from $T_0$ to $T_0 + \Delta T$.
To first order in $\delta \Delta G$,

\begin{equation}    |\Delta T|
    =
    \frac{|\delta \Delta G|}
    {\left| d\Delta G/dT \right|_{T_0}}
    =
    \frac{|\delta \Delta G|}{|\Delta S|}\end{equation}

where $\Delta S$ is the entropy difference between the two phases at the transition.

For the transitions studied here, the entropy difference at the transition is about $1\,k_B$/atom for melting of the pure elements and about $0.5\,k_B$/atom for the order-disorder transition at $x_\mathrm{Cu}=0.5$.
An error of 1~meV/atom in $\Delta G$ therefore shifts the melting temperatures by about 12~K and the order-disorder transition temperature by about 23~K.
For melting, we tested this relation against the melting temperatures obtained by recomputing the free energy crossings with the FEP corrections (Methods Sec.~4.6).
The estimates to first order agree with the recomputed shifts of 15--44~K to within about 20\%.

The statistical error of the switching runs is at most 0.5~meV/atom (Supplementary Note~\ref{si:diss}).
Its influence on the phase boundaries is therefore 10--20~K in the center of the diagram and much smaller toward the pure elements.

\section{Fitting Errors}

\begin{figure*}
    \includegraphics[width=\textwidth]{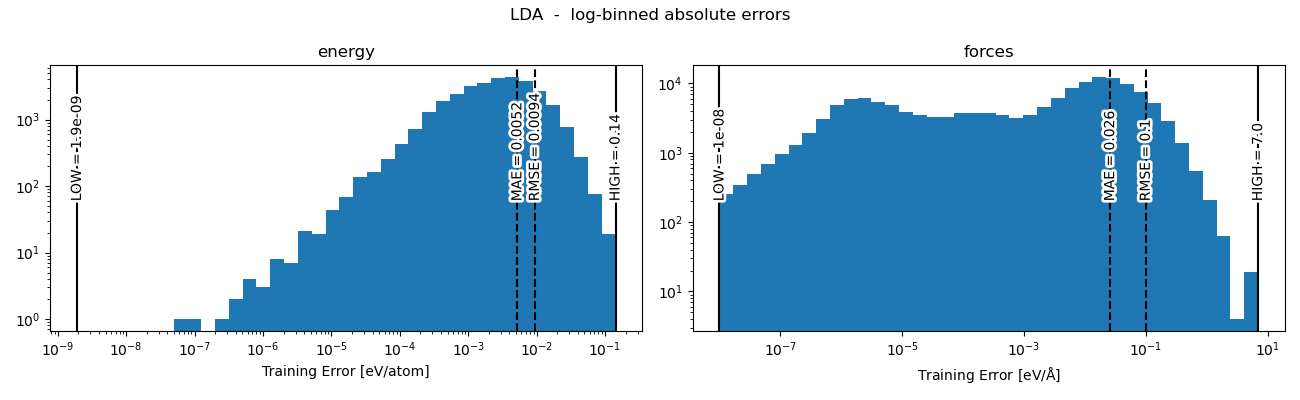}
    \includegraphics[width=\textwidth]{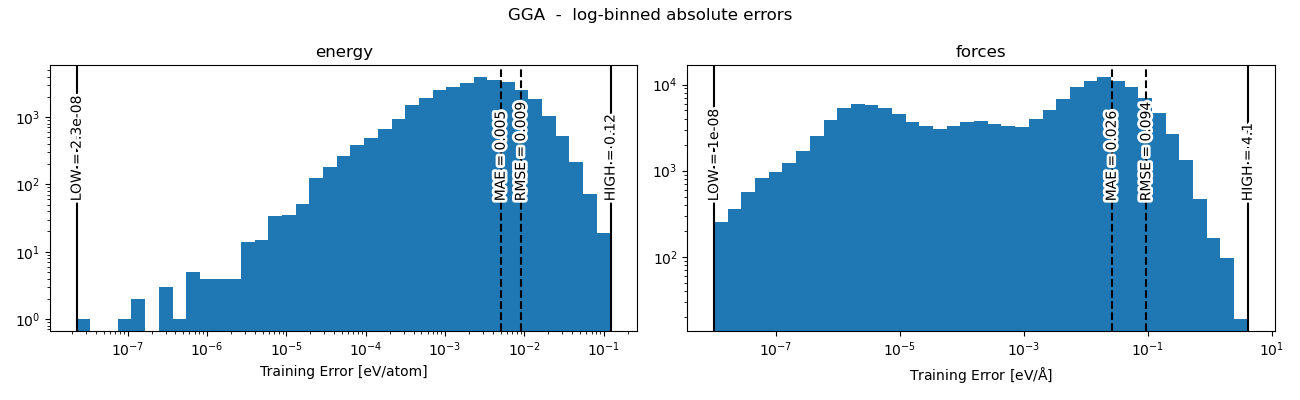}
    \includegraphics[width=\textwidth]{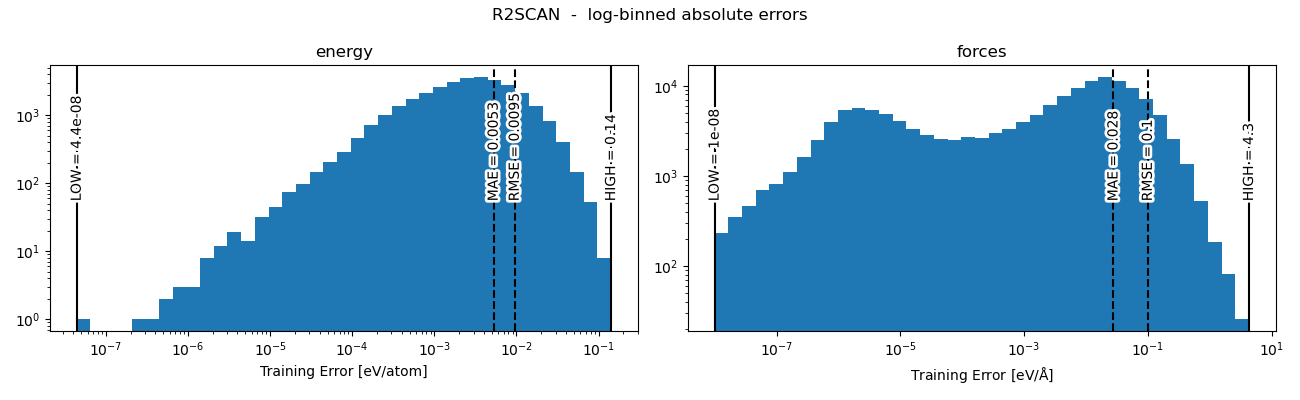}
    \caption{\label{fig:trainerror}\
        Histogram of energy and force errors on the training set for all three functionals, LDA (top), PBE (middle), and r$^2$SCAN (bottom).
        Counts are given on a log scale and within logarithmic bins to resolve more detail.
        Vertical lines show minimum, mean absolute, root mean square, maximum errors.
    }
\end{figure*}

All three ACE potentials fit the training set very well, with mean absolute errors (MAE) of $5\;\mathrm{m}\,\mathrm{eV}$ and root mean square errors (RMSE) of $9\;\mathrm{m}\,\mathrm{eV}$, across all references.
Force MAE at $\sim 30\;\mathrm{m}\,\mathrm{eV}$ also indicate a good fit.
Fig.\ \ref{fig:trainerror} shows histograms of the energy and force error over all training points.
Due to ASSYST's wide energy range in the training data, RMSE are notably larger than MAE over the whole dataset.
As shown in the next section, low energy structure are better described.

\section{Energy Volume Curves}

\begin{figure*}
    \includegraphics[width=\textwidth]{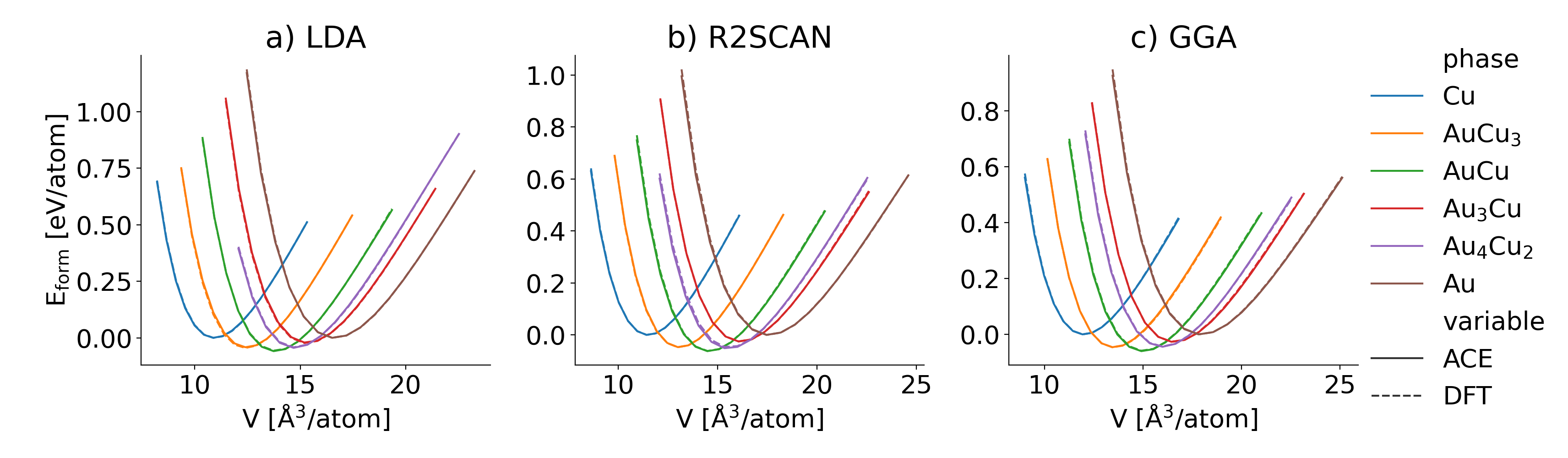}
    \caption{\label{fig:ev}\
        Energy-volume curves against DFT for a) ACE-LDA, b) ACE-PBE, c) ACE-r$^2$SCAN over a
        volumetric strain range -25\,\% to 40\,\%
        for the five stable phases and the experimentally stable Au$_3$Cu.
    }
\end{figure*}

Fig. \ref{fig:ev} shows the energy volume curves for all three ACE potentials presented in the main text.
Over the full range of strains tested, the agreement is excellent, with the energy MAE over all strains of 3 meV/atom, rising up to 20 meV/atom in the compressive regime.
For equilibrium energy and volume, we observe less than 2 meV/atom and $0.02\,\text{\AA}^3$/atom MAE respectively.
Fitting the equations of state with quintic splines, we also find very good agreement of the bulk modulus and its derivative within 2\,\% and 6\,\% respectively.

\section{Elastic Constants}

\begin{figure*}
       \includegraphics[width=\textwidth]{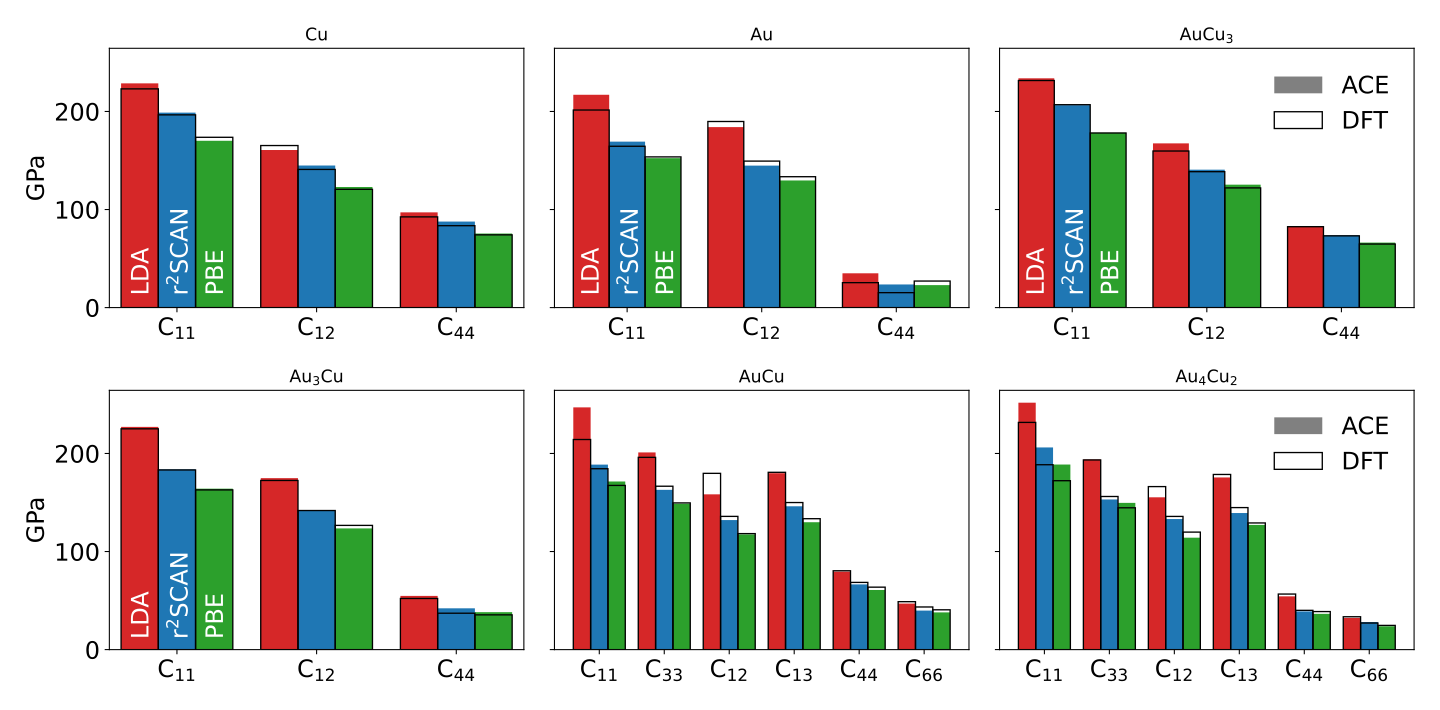}
    \caption{\label{fig:elastic}\
        Elastic constants for LDA (red), r$^2$SCAN (blue), and PBE (green)
        for the five stable phases and the experimentally stable Au$_3$Cu.
        ACE prediction for the independent constants (x-axis) are given as 
    }
\end{figure*}

We calculated elastic tensors using \texttt{elastic}\cite{pawel_t_jochym_2018_1254570}.
Fig. \ref{fig:elastic} shows the independent components by phase as bar plots.
For all three ACE potentials, the agreement is good, although LDA is slightly worse described at 7 GPa MAE compared to 3 GPa MAE for PBE and 4 GPa for r$^2$SCAN potentials.
LDAs larger deviation is largely due to overestimating $C_{11}$ and softening $C_{12}$ in Au, AuCu, and Au$_4$Cu$_2$.

\section{Phonon Density of States}

\begin{figure*}
    \includegraphics[width=\textwidth]{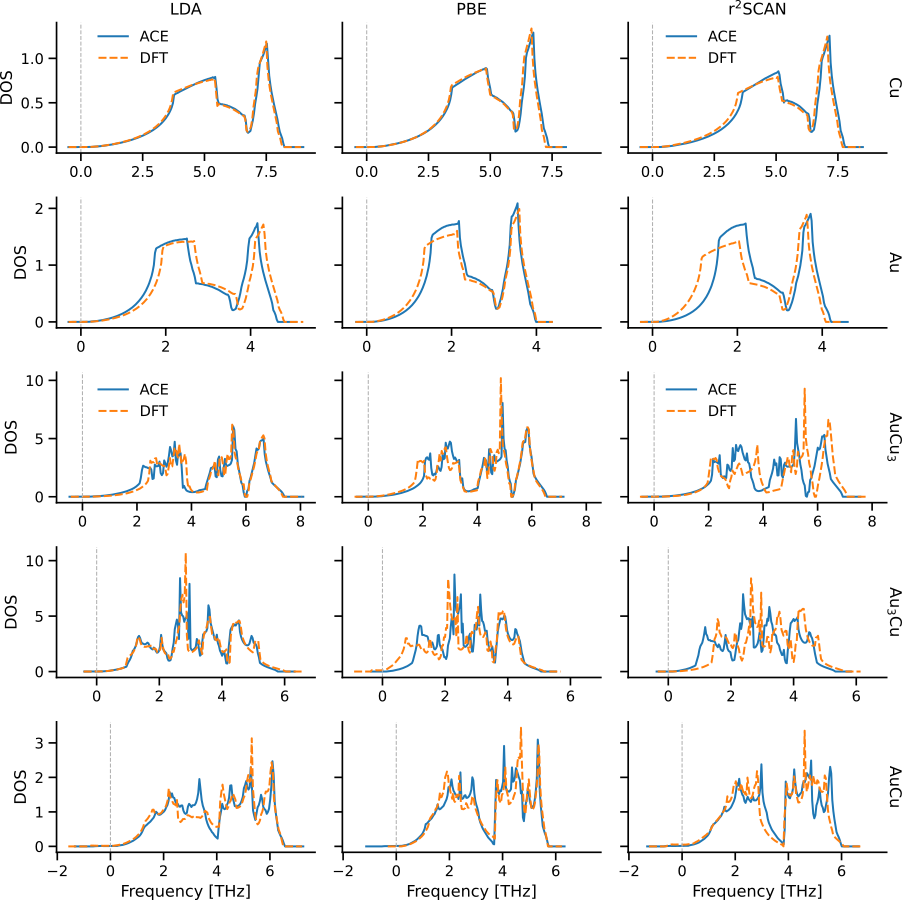}
    \caption{\label{fig:phonons}
        Phonon spectra for LDA, PBE, and r$^2$SCAN (left to right) descriptions of
        Cu, Au, AuCu$_3$, Au$_3$Cu, AuCu (top to bottom).
        Orange dashed lines show the DFT predictions, blue solid lines the ACE predictions.
    }
\end{figure*}

Finally, we compare the density of states (DOS) of phonons for all three models in fig. \ref{fig:phonons}.
Qualitative agreement is observed throughout, with some frequency shifts particularly noticable in Au and AuCu$_3$.
The calculations were performed using \texttt{phonopy}.\cite{phonopy-phono3py-JPSJ}

\section{Convergence of alchemical transformations}

\begin{figure*}
    \includegraphics[width=\textwidth]{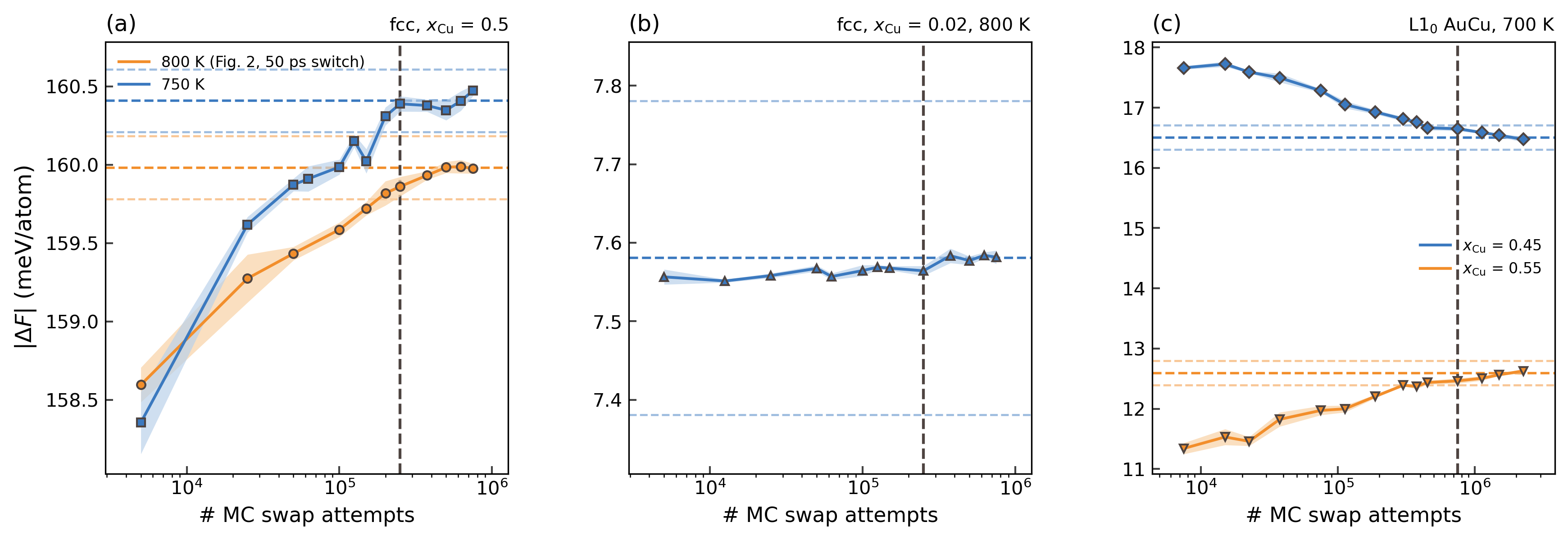}
    \caption{\label{fig:convergence_states}
        Convergence of the alchemical free energy difference with the number of MC identity exchange attempts.
        (a) FCC solid solution at $x_{\mathrm{Cu}}=0.5$ at 750 and 800~K.
        (b) Dilute FCC solid solution at $x_{\mathrm{Cu}}=0.02$ and 800~K.
        (c) Off-stoichiometric L1$_0$ AuCu at $x_{\mathrm{Cu}}=0.45$ and $0.55$ at 700~K.
        The dark horizontal dashed lines indicate the converged values and the lighter dashed lines indicate a $\pm 0.2$~meV/atom interval.
        The vertical dashed lines indicate the number of swap attempts used in the corresponding production calculations.
    }
\end{figure*}

In addition to the convergence analysis for the FCC solid solution at $x_{\mathrm{Cu}}=0.5$ in the main text, we repeat the convergence analysis for additional compositions and phases.
We consider the equiatomic FCC solid solution at 750~K, close to the predicted order-disorder transition, a dilute FCC solid solution at $x_{\mathrm{Cu}}=0.02$ and 800~K, and off-stoichiometric L1$_0$ AuCu at $x_{\mathrm{Cu}}=0.45$ and $0.55$ at 700~K.

Fig.~\ref{fig:convergence_states} shows the calculated free energy difference as a function of the total number of MC identity exchange attempts.
For the equiatomic FCC solid solution, the free energy at 750~K enters the $\pm0.2$~meV/atom interval at about $2\times10^5$ attempts, similar to the 800~K calculation in the main text, and lies 0.02~meV/atom from the converged value at the production sampling.
The dilute FCC solid solution depends little on the number of swap attempts and lies 0.02~meV/atom from the converged value at the production sampling.
The calculations for L1$_0$ away from stoichiometry converge most slowly.
In these calculations the atoms form three groups, the two sublattices and the atoms being transformed, and swaps are attempted between each pair of groups, so that the production calculations use 750\,000 attempts per switching direction.
At this sampling both compositions lie within 0.14~meV/atom of the converged value and remain within the interval as the number of attempts is increased further.
With the sensitivity derived in Supplementary Note~4, the largest of these offsets corresponds to about 3~K in the transition temperature.
In the liquid, the arrangement of the species is not sampled with identity exchange moves but evolves by atomic diffusion during the molecular dynamics.

Finally, we tested the convergence of the FCC solid solution free energy with system size at $x_{\mathrm{Cu}}=0.5$ and 800~K.
For the size convergence test, we used cubic FCC cells containing 500 to 16384 atoms, corresponding to $5^3$ to $16^3$ conventional unit cells.
Three independent calculations were performed for each size.
The number of identity exchange attempts was scaled with the number of atoms so that the sampling effort per atom remained approximately constant for all system sizes.

\begin{figure}[htp!]
    \centering
    \includegraphics[width=1.0\linewidth]{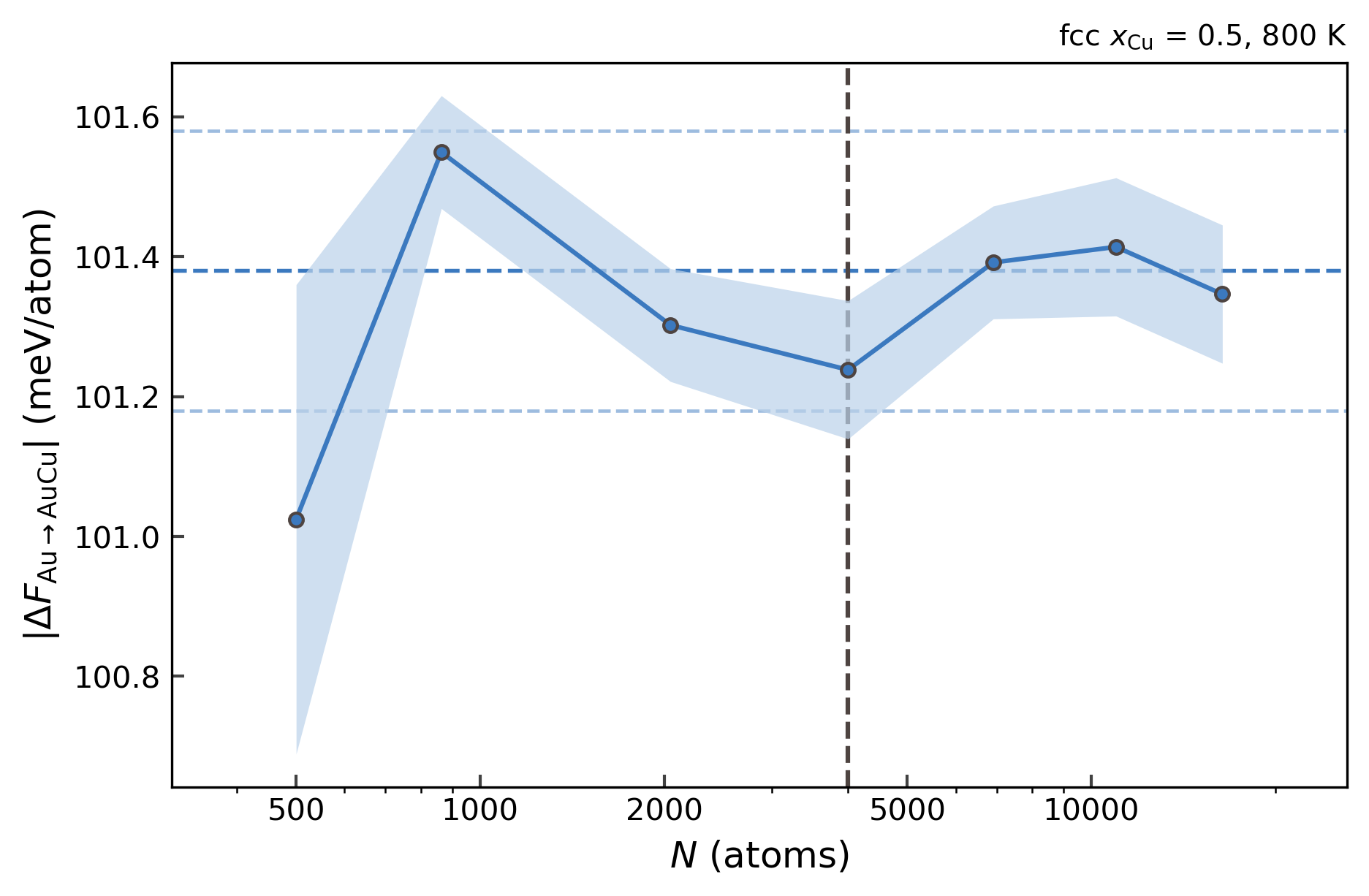}
    \caption{
    Convergence of the alchemical free energy contribution with system size for the FCC solid solution at $x_{\mathrm{Cu}}=0.5$ and 800~K.
    The horizontal dashed line indicates the converged value obtained from the largest cells, and the lighter dashed lines indicate a $\pm0.2$~meV/atom convergence interval.
    The vertical dashed line marks the system size of 4000 atoms used in this work.
    Points show averages over three independent calculations, and the shaded region indicates the standard error estimated from the pooled variance between independent runs.
    }
    \label{fig:size}
\end{figure}

Figure~\ref{fig:size} shows the alchemical contribution, which contains the chemical and configurational contributions associated with the change in composition.
From 864 atoms onward, all calculated values lie within the $\pm0.2$~meV/atom convergence interval defined from the largest cells.
No systematic dependence on system size is resolved within the statistical precision of the calculations.
The cell containing 500 atoms lies outside this interval and also shows substantially larger variation between independent runs.
The system size used in this work, containing 4000 atoms, is therefore within the converged regime for the alchemical contribution.
We restrict the size study to the FCC solid solution, the phase in which the chemical disorder is sampled and could be limited by the cell size.
The stoichiometric ordered phases are computed without identity exchange moves, and their chemical disorder is limited to dilute thermal antisites.
Their remaining size dependence is therefore dominated by the vibrational $1/N$ term of the reference free energy, which largely cancels between phases of similar size.

\section{AuCu II}
\label{si:aucuII}

Experimentally, L1$_0$ AuCu transforms at about 658~K into AuCu II, an orthorhombic superstructure in which the L1$_0$ ordering is shifted by half a lattice vector every five unit cells along one axis, and AuCu II disorders at about 683~K \cite{Bonneaux1999,Battezzati2001,Anraku2009}.
To test whether AuCu II is stable with the present potentials, we computed its free energy with ACE-LDA.

\begin{figure}[htp!]
    \centering
    \includegraphics[width=1.0\linewidth]{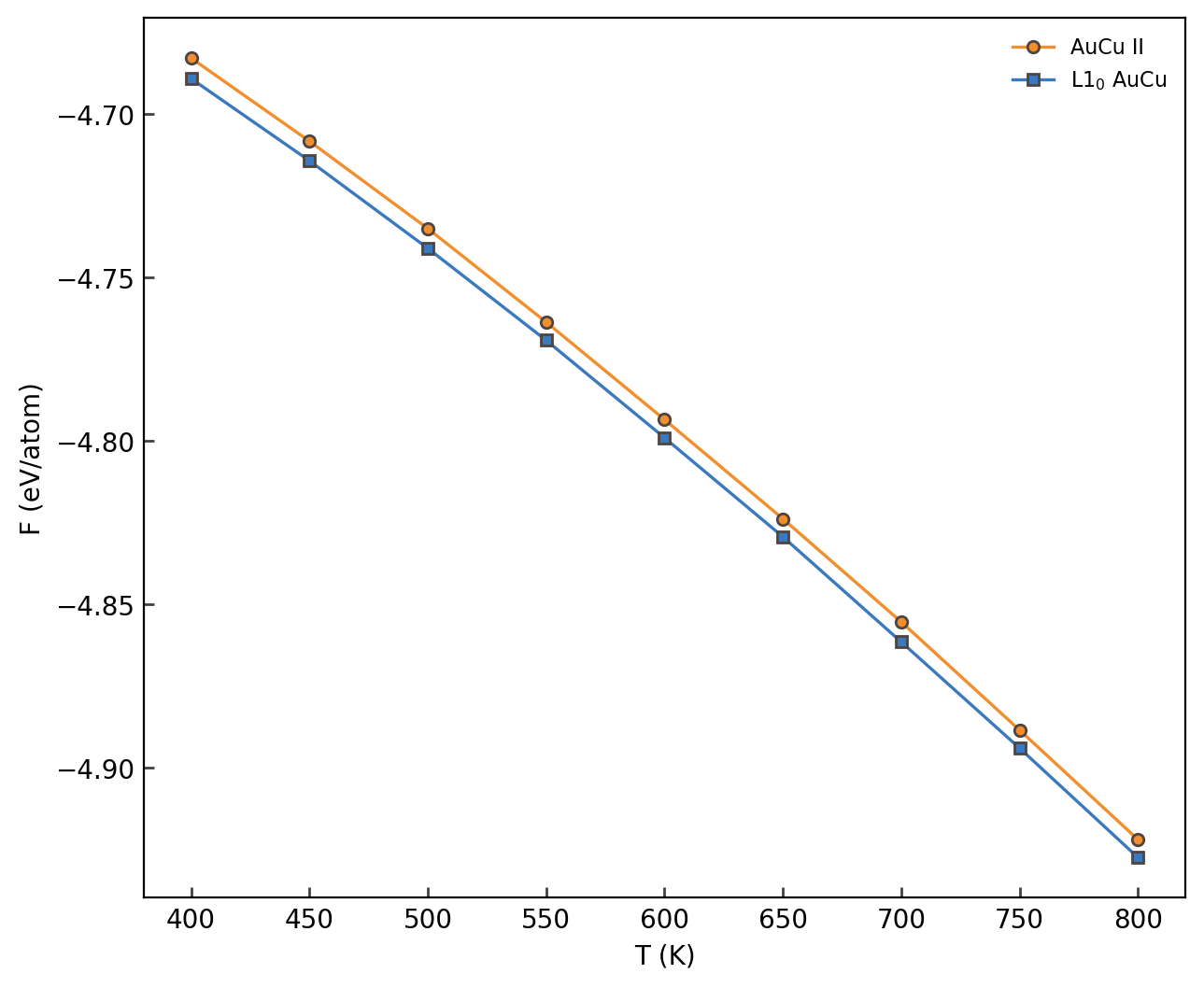}
    \caption{Free energies of AuCu II and L1$_0$ AuCu as a function of temperature, computed with ACE-LDA.}
    \label{fig:aucuII}
\end{figure}

We constructed the AuCu II cell with a period of five L1$_0$ unit cells and replicated it to a supercell of 3920 atoms.
Its free energy was computed between 400 and 800~K with the same Frenkel--Ladd procedure as for the other ordered phases, with all three cell lengths relaxed at each temperature.
Table~\ref{tab:aucuII} and Fig.~\ref{fig:aucuII} compare it with L1$_0$ AuCu.

\begin{table*}[htp!]
\centering
\caption{Free energy of AuCu II relative to L1$_0$ AuCu with ACE-LDA, in meV/atom.}
\label{tab:aucuII}
\begin{tabular}{lccccccccc}
\toprule
$T$ (K) & 400 & 450 & 500 & 550 & 600 & 650 & 700 & 750 & 800 \\
\midrule
$F_{\mathrm{AuCu\,II}} - F_{\mathrm{L1}_0}$ & 6.3 & 6.0 & 6.1 & 5.6 & 5.6 & 5.5 & 6.1 & 5.5 & 5.5 \\
\bottomrule
\end{tabular}
\end{table*}

AuCu II lies 5.5--6.3~meV/atom above L1$_0$ AuCu at all temperatures between 400 and 800~K, nearly independent of temperature, and therefore does not become stable.
The same holds at 0~K for all three ACE potentials.
After full relaxation of the atomic positions and the cell, AuCu II lies 8.2, 4.8 and 6.7~meV/atom above L1$_0$ AuCu for ACE-LDA, ACE-PBE and ACE-r$^2$SCAN, respectively.
This is consistent with the Materials Project entry for this structure (mp-582681, PBE), which lies 10~meV/atom above the convex hull \cite{Jain2013}.
AuCu II therefore does not enter any of the computed phase diagrams.
Experimentally, AuCu II is stable over a window of about 25~K, so the present potentials do not reproduce its stability.
Superstructures with a long period depend on small energy differences between ordered stackings, which are difficult to describe with semilocal functionals.

\bibliographystyle{simplejournal}
\bibliography{references}